\documentclass[aps,prd,preprint,amsmath,citeautoscript,longbibliography,nofootinbib]{revtex4-1}
\usepackage[utf8]{inputenc}
\usepackage{bm}
\usepackage{amsmath,mathrsfs,amsfonts}
\usepackage{graphicx,afterpage}
\usepackage{subcaption}
\usepackage{amsmath,latexsym}
\usepackage{color,soul}
\usepackage{float}
\usepackage[section]{placeins}
\usepackage{silence}
\begin{document}
\title{Generalized Hidden Conformal Symmetry in {\textcolor{black}{Quadratic}} $f(T)$ Gravity}
\author{Bardia H. Fahim {\footnote{bardia.fahim@usask.ca}}, A. M. Ghezelbash{\footnote{masoud.ghezelbash@usask.ca}}}

\affiliation{Department of Physics and Engineering Physics, University of Saskatchewan, Saskatoon SK S7N 5E2, Canada}
\date{\today}
\begin{abstract}
We find the non-extremal charged rotating black holes in {\textcolor{black}{quadratic}} $f(T)$ gravity are holographically dual to two different hidden conformal field theories.  The  two  conformal field theories can be merged to find a very general hidden conformal field theory, which is generated  by the $SL(2,\mathbb{Z})$ modular group. We also carry out the the calculation to the extremal limit of the black holes, and find the corresponding dual quantities. Contrary to the existence of two different dual conformal field theories for the extremal charged rotating black holes in Einstein gravity, we find only one dual theory exists for the extremal charged rotating black holes in  {\textcolor{black}{quadratic}} $f(T)$ gravity.
\end{abstract}
\maketitle


\section{Introduction}
\label{sec:intro}

The black hole holography states the correspondence between the rotating black holes and the conformal field theory (CFT). The correspondence has been extensively constructed for the black holes in the Einstein theory of gravity.  In fact, the first of such attempts was carried out in   \cite{stro}, where the authors showed  the correspondence between the extremal Kerr black holes and the associated CFT. 

The results of the correspondence relate not only the microscopic Bekenstein-Hawking entropy, but also the super-radiant modes of the extremal Kerr black holes  to the dual quantities 
to a conformal field theory.  The diffeomorphisms of the near-horizon geometry for the Kerr black holes generate  the generators of the  conformal symmetry. The  Kerr/CFT correspondence  later was extended for the different four and higher dimensional extremal rotating black holes in Einstein gravity \cite{KerrCFT1}-\cite{SaktiAnnPhys2020}. A common feature of near horizon geometry for all the different extremal rotating black holes is that 
the near horizon geometry possesses a copy of the AdS space. In fact, the enhancement of the AdS isometries makes the Virasoro algebra for a dual conformal field theory to the black holes.

Though the correspondence for the extremal rotating black holes is quite interesting due to AdS isometries of the near horizon geometry, however for the  non extremal rotating black holes, the absence of the AdS isometries makes the situation more complicated.  
Instead of  usual procedure for the extremal rotating black holes, we may find the dual conformal symmetry (known as the hidden conformal symmetry)  for the non extremal rotating black holes, by searching in the solution space of  a probe field in the background of non extremal rotating black holes \cite{cms}.
The hidden CFT has been  constructed for different types of the non extremal rotating black holes in four and higher dimensional Einstein gravity \cite{othe}-\cite{KSChen}.

The conformal structure dual to the  charged rotating black holes is even richer than non-charged rotating black holes. In fact  for some charged rotating black holes, there  are two separate dual  CFTs.  As an example, for the four-dimensional Kerr-Newman black holes, we can find two separate dual CFTs. The two dual CFTs are dual to the physical rotational parameter and the electric charge of the black holes \cite{Chen-JQpic}. 
However the existence of two dual CFTs depends explicitly on the black holes, and not all charged rotating black holes enjoy having two dual CFTs. As an example,  the charged rotating Kerr-Sen black holes are dual only to one CFT, corresponding to the rotational parameter of the black holes  \cite{sen, Ghezelbash}. 

In fact,  there is no dual CFT to the physical electric charge of the Kerr-Sen black holes, as   
the non-gravitational fields in the Kerr-Sen solutions, don’t contribute to the central charge of the dual CFT \cite{GH2}.  {\textcolor{black}{The duality between the non-extremal rotating black holes (as well as extremal black holes) and the CFT have been shown to be valid through comparison between the macroscopic black hole quantities, as solutions of the  General Relativity (GR), and the microscopic CFT quantities. In particular,  in the context of duality, there is a perfect match between the macroscopic Bekenstein-Hawking entropy of the rotating black holes and the entropy of the CFT which is computed by the Cardy formula. Another very interesting result which supports the duality is coming from the study of the super-radiant scattering off the rotating black holes. It was shown that the bulk scattering amplitudes are in precise agreement with the scattering results from CFT.  The scattering amplitudes of CFT are completely fixed by the conformal invariance.
}}

In this article, our focus is on the charged rotating black holes in the context of  modified theories of gravity  (MTG) {\textcolor{black} {\cite{24}-\cite{56}, and in particular, the quadratic $f(T)$ gravity.}} The MTG  have been known for a long time, are theories to address the issues such as dark energy, since the accelerating expansion of the universe was discovered. These theories are of great importance in cosmology, to handle the problems, for which general relativity does not provide any solutions. Moreover, in the appropriate limits, these theories should produce the same results as general relativity. 

{\textcolor{black}{In modified theory of $f(T)$ gravity, we use the curvature-free connection, instead of the usual torsion-free Levi-Civita connection in general relativity. The simplest possibility is to use the torsion scalar $T$, instead of Ricci scalar in the action of general relativity. The theory is called Teleparallel Equivalent of General Relativity (TEGR). 
The TEGR is completely equivalent to GR, as a calculation shows that the action of TEGR is equal to GR action up to a total derivative \cite{OLD6}.
We can consider a more general theory, where the action is a function of torsion scalar $f(T)$, which leads to a new class of MTG \cite{42}-\cite{51}. }}

Inspired with the possible existence of two dual CFTs to the charged rotating black holes of GR, in this article, we consider a charged rotating black hole in quadratic $f(T)$ theory of gravity. {\textcolor{black}{Our interest in the class of $f(T)$ gravity is that in these theories unlike GR (as well as $f(R)$ gravity), the building block of the theory is the torsion. The existence of torsion may or may not hinder the existence of the duality between the rotating black holes of $f(T)$ and CFT. Of course to simplify the tedious calculation, we consider the simplest class of $f(T)$ gravity. Hence in this article, we focus only on the quadratic $f(T)$ gravity, where $f(T)=T+\alpha T^2$, where $T$ is the scalar torsion and $\alpha$ is a dimensional negative parameter. Moreover for this class of $f(T)$ gravity, we can use the well known four-dimensional charged rotating black hole \cite{OLD3} to establish or rule out the duality between the black hole and CFT. Of course, establishing the duality for more general form of $f(T)$ gravity, requires the existence of rotating black holes, and the feasibility of solving the field equations.}}

We organize the article as follows. In section \ref{sec:sec2scalarfield}, we review briefly the $f(T)$ theory which is coupled to the cosmological constant and the Maxwell fields.  We also consider the Klein-Gordon equation for a probe scalar field in the background of a charged rotating  {\textcolor{black}{quadratic}} $f(T)$ black hole.
In section \ref{sec:sec3hidden}, we explore and find the hidden conformal symmetry  for  the charged rotating  {\textcolor{black}{quadratic}} $f(T)$  black holes. We explicitly construct the two different CFTs which are dual to the rotational parameter and the electric charge of the black holes. 
In section \ref{sec:sec4:absorption}, we consider the  extremal charged rotating  {\textcolor{black}{quadratic}} $f(T)$ and find that there is only one consistent CFT dual to the extremal black holes.
We wrap up the article by concluding remarks in section \ref{sec:concl}.

\section{Charged Scalar Field in the Background of Charged Rotating Black Holes in quadratic $f(T)$ Gravity  }\label{sec:sec2scalarfield}

One of the alternative approaches to the theory of general relativity, is the teleparallel gravity, or $f(T)$ gravity. In this theory, teleparallel gravity attributes gravitation to torsion \cite{OLD1}. The basic variables in teleparallel gravity are tetrad fields $e_a^{\ \mu}$ with the following relations $e_a^{\ \mu}e^b_{\ \mu}=\delta_a^b$ and $e_a^{\ \mu}e^a_{\ \nu}=\delta^{\mu}_{\nu}$. The metric of the spacetime is related to the tetrads as \cite{OLD2}
\begin{equation}
   g_{\mu\nu}= \eta_{ab} e^a_{\ \mu}e^b_{\ \nu},
\end{equation}
where $\eta_{ab}$ is the metric of the four-dimensional Minkowski spacetime, and $a$ and $b$ are the internal space indices $a=(0,1,2,3)$. 
{\textcolor{black}{
Moreover, the curvature free Weitzenbock connection is defined as
\begin{equation}
    W^{\alpha}_{\mu\nu}=e_a \ ^{\alpha}\partial_{\nu}e^a \ _{\mu}+e_a \ ^{\alpha}\omega^a \ _{b\nu}e^b \ _{\mu},\label{WW}
\end{equation}
where $\omega^a \ _{b\nu}$ is the spin-connection. The non-vanishing torsin of the Weitzenbock connection is given as 
\begin{equation}
    T^{\alpha}_{\ \mu\nu}=W^{\alpha}_{\ \nu\mu}-W^{\alpha}_{\ \mu\nu}.
\end{equation}
We note that since the spin-connection appears as a surface term in the action of the TEGR \cite{OLD6}, the choice of the spin-connection does not affect the equations of motion. This is not true in $f(T)$ gravity, where the Lagrangian is non-linear in torsion scalar. Therefore, the solution of the field equations in $f(T)$ gravity depends on the frame, which would not be invariant under the local Lorenz transformation \cite{N4}-\cite{N6}. This is because the field equations represent both inertia and gravity. As there is not an agreement on a common opinion of the covariant approach to the teleparallel gravity, we follow the approach proposed in \cite{N3}. Under a Lorentz transformation, the spin-connection $\omega^c \ _{d\mu}$ transforms as
\begin{equation}
    \omega'^a \ _{b\mu}=\Lambda^a \ _c \omega^c \ _{d\mu} \Lambda_b \ ^d+\Lambda^a \ _c\partial_{\mu}\Lambda_b \ ^c.
\end{equation}
Since the spin-connection of teleparallel gravity is purely inertial connection, there exists a class of frame in which the spin-connection vanishes $\omega^a \ _{b\mu}=0$ \cite{N1,N5}, and so the Weizenbock connection in equation (\ref{WW}) simply reduces to $W^{\alpha}_{\mu\nu}=e_a \ ^{\alpha}\partial_{\nu}e^a \ _{\mu}$. We adopt the corresponding proper tetrads to this choice. It is important to realize when one change the frame from one to another, in order to preserve the Lorentz covariance, it is required to apply the transformation on both the tetrad and the spin-connection. Under a local Lorentz transformation, which we denote by $\Lambda^a \ _b$, the tetrad and spin-connection transform simultaneously as
\begin{equation}
    e'^a \ _{\mu}=\Lambda^a \ _b e^b \ _{\mu},
\end{equation}
\begin{equation}
    \omega'^a \ _{b\mu}=\Lambda^a \ _c \omega^c \ _{d\mu} \Lambda_b \ ^d+\Lambda^a \ _c\partial_{\mu}\Lambda_b \ ^c.
\end{equation}
We follow the same approach as \cite{N2}-\cite{N8}, and use the frame in which the spin-connection is zero.
}}

The teleparallel torsion scalar is constructed as 
\begin{equation}
T=T^{\alpha}_{\ \mu\nu}S_{\alpha}^{\ \mu\nu},
\end{equation}
 where $S_{\alpha}^{\ \mu\nu}$ is the superpotential tensor 
 \begin{equation}
 S_{\alpha}^{\ \mu\nu}=\frac{1}{2}(K^{\mu \nu}_{\  \alpha}+\delta^{\mu}_{\alpha}T^{\beta\nu}_{\ \beta}-\delta^{\nu}_{\alpha}T^{\beta\mu}_{\ \beta}),
 \end{equation}
 and $K^{\mu \nu}_{\  \alpha}$ is the contortion tensor \cite{OLD3}.

In this paper, we consider a four-dimensional charged rotating AdS black hole in {\textcolor{black}{quadratic}} $f(T)$-Maxwell theory, where the action for an asymptotically AdS spacetime, is given by
\begin{equation}
    S=\frac{1}{2\mathcal{G}}\int d^4x \lvert e \rvert(f(T)-2\Lambda-F\wedge ^*F). \label{e1}
\end{equation}
In action (\ref{e1}), $e = \det(e_a^\mu) $, and  we consider the class of  $f(T)$ theories, which $f(T)$ is given by
\begin{equation}
    f(T)=T+\alpha T^2,\label{fTa}
\end{equation}
where $T$ is the scalar torsion and $\alpha$ is the dimensional negative parameter. Moreover, $\mathcal{G}=2\Omega_2G_4$, where $\Omega_2$ is the volume of the 2-sphere and $G_4=1$ is the four-dimensional Newton's gravitational constant. 
{\textcolor{black}{In this paper, we use the Planck units, in which $c=G_4= \hbar=k_B =1$.
}}
The cosmological constants is related to the length scale of the AdS spacetime $l$ as
\begin{equation}
    \Lambda=-3/l^2.
\end{equation}
The line element of the four-dimensional charged rotating  AdS black hole in {\textcolor{black}{quadratic}} $f(T)$-Maxwell theory is given by \cite{OLD3}
\begin{equation}
    ds^2=-A(r)(\Xi dt- \Omega d\phi)^2+\frac{dr^2}{B(r)}+r^2\frac{(\Omega dt-\Xi l^2 d\phi)^2}{l^4}+\frac{r^2}{l^2}dz^2, \label{lineel}
\end{equation}
where $\Omega$ is  the rotation parameter,  and $\Xi=\sqrt{1+\Omega^2/l^2}$. The metric functions $A(r)$ and $B(r)$ are given by \cite{OLD4}
\begin{equation}
    A(r)=r^2\Lambda_{eff}-\frac{M}{r}+\frac{3Q^2}{2r^2}+\frac{2Q^3\sqrt{6|\alpha|}}{6r^4}, \label{metra}
\end{equation}
\begin{equation}
    B(r)=A(r)\beta(r),\label{metrb}
\end{equation}
where
\begin{equation}
    \beta=(1+Q\frac{\sqrt{6|\alpha|}}{r^2})^{-2},\label{beta}
\end{equation}
\begin{equation}
    \Lambda_{eff}=\frac{1}{36|\alpha|}. \label{Lameff}
\end{equation}
In equations (\ref{metra})-(\ref{beta}), $Q$ and $M$ are the the charge and the mass of the black hole, respectively.
{\textcolor{black}{Equation ({\ref{Lameff}}) shows that $\alpha$ cannot be zero, otherwise the effective cosmological constant $\Lambda_{eff}$ and the metric functions $A(r)$ and $B(r)$, diverge. We note that the metric (\ref{lineel}) in the limit $\Omega=0$ reduces to the static charged black hole configuration, as studied in \cite{OLD5}.}} The scalar torsion is given by \cite{OLD3}
\begin{equation}
    T(r)=\frac{4A'(r)B(r)}{rA(r)}+\frac{2B(r)}{r^2},
\end{equation}
where prime denotes the derivative with respect to the coordinate $r$.
The electromagnetic potential is
\begin{equation}
    A_{\mu}dx^{\mu}=-(\frac{Q}{r}+\frac{Q^2\sqrt{6|\alpha|}}{3r^3})[\Omega d\phi-\Xi dt]. \label{emp}
\end{equation}

In order to explore the hidden conformal symmetry, we consider a massless charged scalar field $\Phi$ with charge $q$, as a probe in the background of the black hole (\ref{lineel}). The non-minimally coupled Klein-Gordon equation for the massless charged scalar field $\Phi$ is
\begin{equation}
    \left( {\nabla _\mu   - iqA_\mu  } \right)\left( {\nabla ^\mu   - iqA^\mu  } \right)\Phi  = 0,\label{eq1}
\end{equation}
where $A_\mu$ is given by (\ref{emp}). As we are working in the $f(T)$ frame, the covariant derivative $\nabla_{\mu}$ in the Klein-Gordon equation (\ref{eq1})  where applied on an arbitrary vector $v^{\nu}$, is given by the following equation,
\begin{equation}
    \nabla_{\mu}v^{\nu}=\partial_{\mu}v^{\nu}+\Gamma^{\nu}_{\ \rho \mu}v^{\rho},
\end{equation}
The connection $\Gamma^{\nu}_{\ \rho \mu}$ is constructed as \cite{OLD6}
\begin{equation}
    \Gamma^{\nu}_{\ \rho \mu}=\mathring{\Gamma}^{\nu}_{\ \rho \mu}+K^{\nu}_{\ \rho \mu}, 
\end{equation}
where $\mathring{\Gamma}^{\nu}_{\ \rho \mu}$ is the zero-torsion Christoffel (Levi-Civita) connection, and $K^{\nu}_{\ \rho \mu}$ is the contortion tensor. The contorsion tensor is defined in terms of the torsion tensor by
\begin{equation}
    K^{\nu}_{\ \rho \mu}=\frac{1}{2}(T_{\mu \ \rho}^{\ \nu}+T_{\rho \ \mu}^{\ \nu}-T^{\nu}_{\ \rho \mu}).
\end{equation}
Since the black hole solution (\ref{lineel}) has three Killing vectors, we separate the coordinates in the scalar field as
\begin{equation}
\Phi \left( {t,r,\phi,z } \right) = \exp \left( {- i\omega t+im\phi  +ikz} \right)R\left( r \right).\label{eq2}
\end{equation}

Plugging equation (\ref{eq2}) in the Klein-Gordon equation (\ref{eq1}), we find that the radial function $R(r)$ satisfies
\begin{equation}
B(r)\frac{d^2R(r)}{dr^2}+\frac{6}{rA(r)}\bigg(1/4rB(r)\frac{dA(r)}{dr}+1/12rA(r)\frac{dB(r)}{dr}+A(r)B(r)\bigg)\frac{dR}{dr}+V(r)R(r)=0, \label{rad1}
\end{equation}
where $V(r)$ is given by
\begin{equation}
    V(r)=V_0(r)+qV_1(r)+q^2V_2(r).\label{pot}
\end{equation}
The different terms in (\ref{pot}) are
\begin{eqnarray}
   V_0(r)&=& \frac{-(k^2l^4\Xi^4+((-2\Xi^2k^2+\omega^2)\Omega^2-2\omega m\Xi\Omega+m^2\Xi^2)l^2+k^2\Omega^4)l^2A(r)}{r^2(\Xi^2l^2-\Omega^2)^2A(r)}\nonumber\\
  &+ &  \frac{(\Xi l^2\omega-\Omega m)^2}{A(r)(\Xi^2l^2-\Omega^2)^2},\\
   V_1(r)&=&\frac{2(-\Xi l^2\omega+\Omega m)(Q\sqrt{6|\alpha|}+3r^2)Q}{3r^3A(r)(-\Xi^2l^2+\Omega^2)},\\
    V_2(r)&=&\frac{(2\sqrt{6|\alpha|}Qr^2+3r^4+2Q^2|\alpha|)Q^2}{3r^6A(r)}.
\end{eqnarray}

The positive roots of the metric function $A(r)$ (given by equation (\ref{metra})) denote the location of the horizons of the black hole. We show the outer horizon by $r_+$. In the near-horizon region, we expand the metric function $A(r)$,
as a quadratic polynomial in $(r-r_+)$, such as 
\begin{equation}
    A(r) \simeq K\left( {r - {r_ + }} \right)\left( {r - {r_ * }} \right), \label{aexp}
\end{equation}
where
\begin{equation}
    K= 15{\Lambda _{eff}} - \frac{3M}{r_+^3} + \frac{3Q^2}{2r_+^4},
\end{equation}
\begin{equation}
    {r_ * } = {r_ + } - \frac{{2{r_ + }\left( {2{r_ + }^4{\Lambda _{eff}} - M{r_ + } + {Q^2}} \right)}}{{10{r_ + }^4{\Lambda _{eff}} - 2M{r_ + } + {Q^2}}}.
\end{equation}
It is worth noting that approximating $A(r)$ by a quadratic polynomial of $(r-r_+)$ as in equation (\ref{aexp}) is necessary to match the radial equation to the Casimir operator of a CFT, hence establishing the hidden conformal symmetry \cite{OLD7, OLD8}.  We note that ${r_ * }$ is not necessarily any of the black hole horizons. We consider the radial equation (\ref{rad1}) at the near-horizon region, where $\omega r_+ \ll 1$. Moreover, we consider a limit where the outer horizon $r_+$ is very close to ${r_ * }$, which is defined by $\left| {{r_ + } - {r_ * }} \right| \ll {r_ + }$. Considering these approximations, the radial equation (\ref{rad1}) simplifies to
\begin{equation}
    \label{rad2}
\frac{d}{dr}\{\left( {r - {r_ + }} \right)\left( {r - {r_ * }} \right)\frac{d}{dr}R\left( r \right)\}+ \left[ {\left( {\frac{{{r_ + } - {r_ * }}}{{r - {r_ + }}}} \right)\mathcal{A} + \left( {\frac{{{r_ + } - {r_ * }}}{{r - {r_ * }}}} \right)\mathcal{B} + \mathcal{C}} \right]R\left( r \right) = 0.
\end{equation}
In (\ref{rad2}), the constant $\mathcal{A} $   is given by
\begin{equation}
    \mathcal{A}=A_1+qA_2+q^2A_3+\omega^2A_4+m^2A_5, \label{A1}
\end{equation}
where
\begin{eqnarray}
   A_1&=& \frac{2(Q\sqrt{6|\alpha|}+r_+^2)^2l^2(K(-1/2k^2l^4\Xi^4+\Omega^2\Xi^2k^2l^2+\Omega\Xi l^2m\omega-1/2k^2\Omega^4)r_*^3}{r_+^5(\Xi^2l^2-\Omega^2)^2r_*^2K^2(r_+-r_*)^2}\nonumber\\
  &+ &  \frac{2(Q\sqrt{6|\alpha|}+r_+^2)^2l^2(-3\Omega\Xi m\omega r_+r_*^2-2\Omega\Xi m\omega r_+^2r_*-\Omega\Xi m\omega r_+^3)}{r_+^5(\Xi^2l^2-\Omega^2)^2r_*^2K^2(r_+-r_*)^2}, \nonumber\\
\end{eqnarray}
\begin{equation}
    A_2=\frac{2(Q\sqrt{6|\alpha|}+r_+^2)^2(r_+^2+2r_+r_*+3r_*^2)(Q\sqrt{6|\alpha|}+3r_+^2)(\Xi l^2\omega-\Omega m)Q}{3r_+^7r_*^2K^2(r_+-r_*)^2(\Xi^2l^2-\Omega^2)},
\end{equation}
\begin{equation}
    A_3=\frac{2(r_+^2+2r_+r_*+3r_*^2)(Q\sqrt{6|\alpha|}+r_+^2)^2Q^2(Q\sqrt{6|\alpha|}r_+^2+3/2r_+^4+Q^2|\alpha|)}{3r_*^2K^2(r_+-r_*)^2r_+^{10}},
\end{equation}
\begin{equation}
    A_4=-\frac{(Q\sqrt{6|\alpha|}+r_+^2)^2l^4(K\Omega^2r_*^3-\Xi^2r_+^3-2\Xi^2r_+^2r_*-3\Xi^2r_+r_*^2)}{r_+^5(\Xi^2l^2-\Omega^2)^2(r_+-r_*)^2K^2r_*^2},
\end{equation}
\begin{equation}
    A_5=-\frac{(Q\sqrt{6|\alpha|}+r_+^2)^2(K\Xi^2l^4r_*^3-\Omega^2r_+^3-2\Omega^2r_+^2r_*-3\Omega^2r_+r_*^2}{r_+^5(\Xi^2l^2-\Omega^2)^2(r_+-r_*)^2K^2r_*^2}.
\end{equation}
Moreover the constant  $\mathcal{B}$ is given by
\begin{equation}
    \mathcal{B}=B_1+qB_2+q^2B_3+\omega^2B_4+m^2B_5, \label{B1}
\end{equation}
where
\begin{eqnarray}
   B_1&=& \frac{2(Q\sqrt{6|\alpha|}+r_+^2)^2l^2(\Omega\Xi m\omega (-3r_+^2-r_*^2)-1/2r_*r_+Kk^2(\Xi^4l^4+\Omega^4))}{r_+^6(\Xi^2l^2-\Omega^2)^2(r_+-r_*)^2K^2}\nonumber\\
  &+ &  -\frac{2(Q\sqrt{6|\alpha|}+r_+^2)^2l^2(r_*r_+(Kk^2l^2\Xi^2\Omega^2+\omega m\Xi(Kl^2-2)\Omega)}{r_+^6(\Xi^2l^2-\Omega^2)^2(r_+-r_*)^2K^2}, \nonumber\\
\end{eqnarray}
\begin{equation}
   B_2=-\frac{2(Q\sqrt{6|\alpha|}+r_+^2)^2(Q\sqrt{6|\alpha|}+3r_+^2)(3r_+^2+2r_+r_*+r_*^2)Q(-\Xi l^2\omega+\Omega m)}{3r_+^9(r_+-r_*)^2K^2(-\Xi^2l^2+\Omega^2)},
\end{equation}
\begin{equation}
     B_3=-\frac{2(Q\sqrt{6|\alpha|}r_+^2+3/2r_+^4+Q^2|\alpha|)(r_+^2+2/3r_+r_*+1/3r_*^2)(Q\sqrt{6|\alpha|}+r_+^2)^2Q^2}{(r_+-r_*)^2K^2r_+^{12}},
\end{equation}
\begin{equation}
    B_4=\frac{(Q\sqrt{6|\alpha|}+r_+^2)^2l^4(-3\Xi^2r_+^2+r_*(K\Omega^2-2\Xi^2)r_+-\Xi^2r_*^2)}{r_+^6K^2(r_+-r_*)^2(\Xi^2l^2-\Omega^2)^2},
\end{equation}
\begin{equation}
    B_5=\frac{(Q\sqrt{6|\alpha|}+r_+^2)^2(-3\Omega^2r_+^2+r_*(\Xi^2Kl^4-2\Omega^2)r_+-\Omega^2r_*^2)}{r_+^6K^2(r_+-r_*)^2(\Xi^2l^2-\Omega^2)^2},
\end{equation}
and  the constant $\mathcal{C} $ is given by
\begin{equation}
    \mathcal{C}=C_1+qC_2+q^2C_3+\omega^2C_4+m^2C_5, \label{C1}
\end{equation}
where
\begin{equation}
    C_1=-\frac{2\omega ml^2\Xi\Omega(r_+^2+r_+r_*+r_*^2)(Q\sqrt{6|\alpha|}+r_+^2)^2}{r_+^6(-\Xi^2l^2+\Omega^2)^2K^2r_*^2},
\end{equation}
\begin{equation}
     C_2=\frac{2(Q\sqrt{6|\alpha|}+r_+^2)^2(r_+^2+r_+r_*+r_*^2)Q(Q\sqrt{6|\alpha|}+3r_+^2)(\Xi l^2\omega-\Omega m)}{3r_+^9(\Xi^2l^2-\Omega^2)K^2r_*^2},
\end{equation}
\begin{equation}
    C_3=\frac{2(Q\sqrt{6|\alpha|}+r_+^2)^2(Q\sqrt{6|\alpha|}r_+^2+3/2r_+^4+Q^2|\alpha|)(r_+^2+r_+r_*+r_*^2)Q^2}{3K^2r_*^2r_+^{12}},
\end{equation}
\begin{equation}
    C_4=\frac{(Q\sqrt{6|\alpha|}+r_+^2)^2(r_+^2+r_+r_*+r_*^2)\Xi^2l^4}{r_+^6r_*^2K^2(\Xi^2l^2-\Omega^2)^2},
\end{equation}
\begin{equation}
    C_5=\frac{(Q\sqrt{6|\alpha|}+r_+^2)^2(r_+^2+r_+r_*+r_*^2)\Omega^2}{r_+^6r_*^2K^2(\Xi^2l^2-\Omega^2)^2}.
\end{equation}

We notice that in the radial equation (\ref{rad2}), the electric charge of the scalar field and the black hole charge are coupled. The presence of the electric charge in the scalar field equation (\ref{eq1}), leads to two different individual CFTs that are holographically dual to the charged rotating AdS black hole in {\textcolor{black}{quadratic}} $f(T)$-Maxwell theory. We call the two different CFTs, simply as  $J$ and $Q$ pictures. In $J$ picture, we assume that the charge of the scalar probe is zero, and we set $q=0$. In $Q$ picture, we assume the zero mode $m=0$, which indicates that the scalar probe co-rotates with the horizon. These pictures have been studied for different black holes, such as Kerr-Newman, Kerr-Sen and Kerr-Newman-NUT-AdS black holes in Einstein frame \cite{OLD9}-\cite{OLD11}. We should note that not all the charged rotating black holes are holographically dual to two conformal pictures \cite{OLD10}.

In order to realize the two dual CFTs to the black hole, we introduce an additional internal degree of freedom $\chi$ for the probe and consider the following expansion \cite{ OLD11}
\begin{equation}
    \Phi=\exp({-i\omega t+im\phi+ikz+iq\chi})R(r). \label{scp}
\end{equation}
 The extra degree of freedom $\chi$ is similar to the $U(1)$ symmetry for the coordinate $\phi$ and is used to uplift the four-dimensional black hole solutions to five-dimensions. We note that this symmetry is equivalent to the original gauge symmetry of the charged rotating black hole in four-dimensions \cite{OLD10}. We apply an $SL(2,\mathbb{Z})$ modular group transformation on $\phi$ and $\chi$ coordinates by combining the two underlying $U(1)$ symmetries, that is associated with $\phi$ and $\chi$ in $J$ and $Q$ picture, respectively. We call this the general picture. The modular group transformation $SL(2,\mathbb{Z})$ that acts on the torus $(\phi,\chi)$ is given by \cite{OLD12}
\begin{equation}
    \binom{\phi'}{\chi'}=\binom{\gamma \ \lambda}{\eta \ \tau}\binom{\phi}{\chi}.
\end{equation}
The modular group transformation does not change the phase of the scalar field (\ref{eq2})
\begin{equation}
    e^{im\phi+iq\chi}=e^{im'\phi'+iq'\chi'},
\end{equation}
which leads to $m=\gamma m'+\eta q'$ and $q=\lambda m'+\tau q'$.
 
 
\section{Hidden Conformal Symmetry for the $f(T)$ Charged Rotating Black Holes}\label{sec:sec3hidden}
In this section, {\textcolor{black}{we find the possible hidden conformal symmetry $SL(2,\mathbb{R})_L\times SL(2,\mathbb{R})_R$ of the near-horizon  scalar field equation (\ref{rad2}) in the background }} of four-dimensional charged rotating AdS black hole in {\textcolor{black}{quadratic}} $f(T)$-Maxwell theory. We define the conformal coordinates $\omega^+, \omega^-$ and $y$ in terms of coordinates $t,r$ and $\phi$ by
\begin{eqnarray}
 \omega^{+}&=&\sqrt{\frac{r-r_{+}}{r-r_{*}}}\exp(2\pi T_{R}\phi +2n_{R}t),
  \label{intro14}\\
  \omega^{-}&=&\sqrt{\frac{r-r_{+}}{r-r_{*}}}\exp(2\pi T_{L}\phi +2n_{L}t),
  \label{intro15}\\
  y&=&\sqrt{\frac{r_{+}-r_{*}}{r-r_{*}}}\exp(\pi( T_{R}+T_{L})\phi +(n_{R}+n_{L})t).
\label{intro16}
 \end{eqnarray} 
 {\textcolor{black}{In equations (\ref{intro14})-(\ref{intro16}), $T_L$ and $T_R$ are the temperatures of the left hand and right hand CFT modes in the Planck units, and $n_L$ and $n_R$ are the mode numbers.
 }}
 We also define the locally conformal operators as
\begin{equation}
    H_1=i\partial_{+} \label{opi},
\end{equation}
\begin{equation}
    H_{-1}=i((\omega^{+})^2\partial_{+}+\omega^{+}y\partial_{y}-y^2\partial_{-}),
\end{equation}
\begin{equation}
    H_0=i(\omega^{+}\partial_{+}+\frac{1}{2}y \partial_{y}),  \label{op3}
\end{equation}
as well as
\begin{equation}
    \overline{H}_1=i\partial_{-},\label{opi2}
\end{equation}
\begin{equation}
    \overline{H}_{-1}=i((\omega^{-})^2\partial_{-}+\omega^{-}y\partial_{y}-y^2\partial_{+}),
\end{equation}
\begin{equation}
    \overline{H}_0=i(\omega^{-}\partial_{-}+\frac{1}{2}y\partial_{y}). \label{opf}
\end{equation}

The operators (\ref{opi})-(\ref{op3}) satisfy the $SL(2,\mathbb{R})_L\times SL(2,\mathbb{R})_R$ algebra
\begin{equation}
 ~~[H_0,H_{\pm1}]=\mp i H_{\pm 1},~~~~~~~~[H_{-1},H_1]=-2iH_0,~~
\label{intro21}
\end{equation}
and similarly for $\overline{H}_1,\overline{H}_0$ and $\overline{H}_{-1}$. 

{\textcolor{black}{We note that $SL(2,\mathbb{R})_L\times SL(2,\mathbb{R})_R$ is a local hidden symmetry, for the solution space of massless scalar field in the near-horizon of the charged rotating AdS black holes (\ref{lineel}), in quadratic $f(T)$-Maxwell theory. The local hidden conformal symmetry $SL(2,\mathbb{R})_L\times SL(2,\mathbb{R})_R$ is generated respectively by the vector fields (\ref{opi})-(\ref{op3}), as well as (\ref{opi2})-(\ref{opf}). Although the coordinate $\phi$ in black hole solution (\ref{lineel}) is periodic with a period of $2\pi$, however the generators (\ref{opi})-(\ref{opf}) are not periodic under the coordinate identification $\phi \sim \phi + 2\pi$. Hence these local symmetries can't be used to produce the global symmetries. In fact, the conformal coordinates (\ref{intro14})-(\ref{intro16}) are identified by $\omega^- \sim e^{4\pi^2T_L}\omega^-$, $\omega^+ \sim e^{4\pi^2T_R}\omega^+$,  and  $y \sim e^{2\pi^2(T_L+T_R)} y$ under the coordinate identification $\phi \sim \phi + 2\pi$. These identifications are the result of subgroup $U(1)_L \times U(1)_R$ element $e^{-4\pi^2iT_L \overline{H}_0}e^{-4\pi^2iT_R {H}_0}$ of the original symmetry group $SL(2,\mathbb{R})_L\times SL(2,\mathbb{R})_R$. 
For a fixed $r$-slice of spacetime, the relations between the conformal coordinates $\omega^{\pm}$ and $(t,\phi)$ look like $\omega^{\pm}=e^{\pm \tau ^\pm}$, where $\tau^-=2\pi T_L\phi+2n_Lt$ and $\tau^+=2\pi T_R\phi+2n_Rt$, up to some $r$-dependence functions. These equations describe exactly the relation between the Minkowski coordinates $(\omega^-,\omega^+)$ and the Rindler coordinates $(\tau^-,\tau^+)$. Hence in the $SL(2,\mathbb{R})_L\times SL(2,\mathbb{R})_R$ invariant Minkowski vacuum state, the observers at fixed Rindler coordinates $(\tau^-,\tau^+)$ observe the Unruh effect \cite{U}, which is a thermal bath of Unruh radiation with a thermal density matrix at temperatures $(T_L,T_R)$ respectively. So, we may conclude that the dual theory to the charged rotating AdS black holes (\ref{lineel}) is a CFT with the finite temperature $(T_L,T_R)$ mixed states. 
}}

We obtain the quadratic Casimir operator, which obeys the $SL(2,\mathbb{R})_L \times SL(2,\mathbb{R})_R$ algebra, from any of two sets of these operators as
 \begin{eqnarray}
\mathcal{H}^2&=& \mathcal{\overline{H}}^2= -H_{0}^2+\frac{1}{2}(H_1H_{-1}+H_{-1}H_{1})\\ \nonumber
&=&\frac{1}{4}
(y^2\partial_{y}^2-y\partial_{y})+y^2\partial_{+}\partial_{-}.
\label{intro28}
\end{eqnarray}
By writing the Casimir operator in terms of the coordinates $(t,r,\phi)$
\begin{eqnarray}
 \mathcal{H}^2&=&(r-r_+)(r-r_*)\partial_r^2+(2r-r_+-r_*)\partial_r+\frac{r_+-r_*}{r-r_*}(\frac{n_L-n_R}{4\pi G}\partial_{\phi}-\frac{T_L-T_R}{4G}\partial_t )^2 \nonumber\\
     &-& \frac{r_+-r_*}{r-r_+}(\frac{n_L+n_R}{4\pi G}\partial_{\phi}-\frac{T_L+T_R}{4G}\partial_t )^2, \label{csm}
\end{eqnarray}
we notice the relation between the radial equation (\ref{rad2}) and the eigenvalue-eigenfunction equation for the quadratic Casimir operator (\ref{csm}), where $G=n_LT_R-n_RT_L$.

First, by considering $q=0$ for the scalar probe in the radial equation (\ref{rad2}), we find the correspondent CFT of the black hole in the $J$ picture. By comparing the Casimir operator (\ref{csm}) and the radial equation (\ref{rad2}), we rewrite the radial equation (\ref{rad2}) in terms of the $SL(2,\mathbb{R})$ quadratic Casimir operator, as
\begin{equation}
    \mathcal{H}^2R(r)={\mathcal{\overline{H}}}^2R(r)=-  \mathcal{C}R(r),
\end{equation}
and we find
\begin{equation}
n_L^J=\frac{f+1}{4(g-h)}\frac{r_+^3K(r_+-r_*)(\Xi^2l^2-\Omega^2)}{(Q\sqrt{6|\alpha|}+r_+^2)l^2}\sqrt{\frac{1}{3\Xi^2r_-^2-r_*(K\Omega^2-2\Xi^2)r_++\Xi^2r_*^2}},\label{nLJ}
\end{equation}
\begin{equation}
    n_R^J=\frac{f-1}{4(g-h)}\frac{r_+^3K(r_+-r_*)(\Xi^2l^2-\Omega^2)}{(Q\sqrt{6|\alpha|}+r_+^2)l^2}\sqrt{\frac{1}{3\Xi^2r_-^2-r_*(K\Omega^2-2\Xi^2)r_++\Xi^2r_*^2}},
\end{equation}
\begin{equation}
    T_L^J=\frac{1}{4\pi(g-h)}\frac{{r_+^3K(r_+-r_*)(\Xi^2l^2-\Omega^2)}}{(Q\sqrt{6|\alpha|}+r_+^2)}\frac{\sqrt{\frac{(K\Omega^2r_*^3+(-r_+^3-2r_*r_+^2-3r_*^2r_+)\Xi^2)r_+}{((-3r_+^2-2r_+r_*-r_*^2)\Xi^2+K\Omega^2r_+r_*)r_*^2}}+1}
    {
    \sqrt{{3\Omega^2r_+^2-r_*(K\Xi^2l^4-2\Omega^2)r_++\Omega^2r_*^2}}
    }
    , \label{TJL}
\end{equation}
\begin{equation}
    T_R^J=\frac{1}{4\pi(g-h)}\frac{{r_+^3K(r_+-r_*)(\Xi^2l^2-\Omega^2)}}{(Q\sqrt{6|\alpha|}+r_+^2)}
    \frac{\sqrt{\frac{(K\Omega^2r_*^3+(-r_+^3-2r_*r_+^2-3r_*^2r_+)\Xi^2)r_+}{((-3r_+^2-2r_+r_*-r_*^2)\Xi^2+K\Omega^2r_+r_*)r_*^2}}-1}
    {
    \sqrt{{3\Omega^2r_+^2-r_*(K\Xi^2l^4-2\Omega^2)r_++\Omega^2r_*^2}}
    }. \label{TJR}
\end{equation}
In equations (\ref{nLJ})-(\ref{TJR}), the quantities $f$,$g$ and $h$ are given by
\begin{equation}
    f=\sqrt{\frac{(K\Xi^2l^4r_*^3+(-r_+^3-2r_*r_+^2-3r_*^2r_+)\Omega^2)r_+}{((-3r_+^2-2r_+r_*-r_*^2)\Omega^2+K\Xi^2l^4r_+r_*)r_*^2}},
\end{equation}
\begin{equation}
    g=\sqrt{-\frac{r_+(K\Omega^2r_*^3+(-r_+^3-2r_*r_+^2-3r_*^2r_+)\Xi^2)}{r_*^2(K\Omega^2r_+r_*+(-3r_+^2-2r_+r_*-r_*^2)\Xi^2)}},
\end{equation}
\begin{equation}
    h=\sqrt{\frac{(K\Xi^2l^4r_*^3+(-r_+^3-2r_*r_+^2-3r_*^2r_+)\Omega^2)r_+}{((-3r_+^2-2r_+r_*-r_*^2)\Omega^2+K\Xi^2l^4r_+r_*)r_*^2}}.
\end{equation}
\begin{figure}[h]
	\centering
	\begin{subfigure}{0.3\linewidth}
		\includegraphics[width=\linewidth]{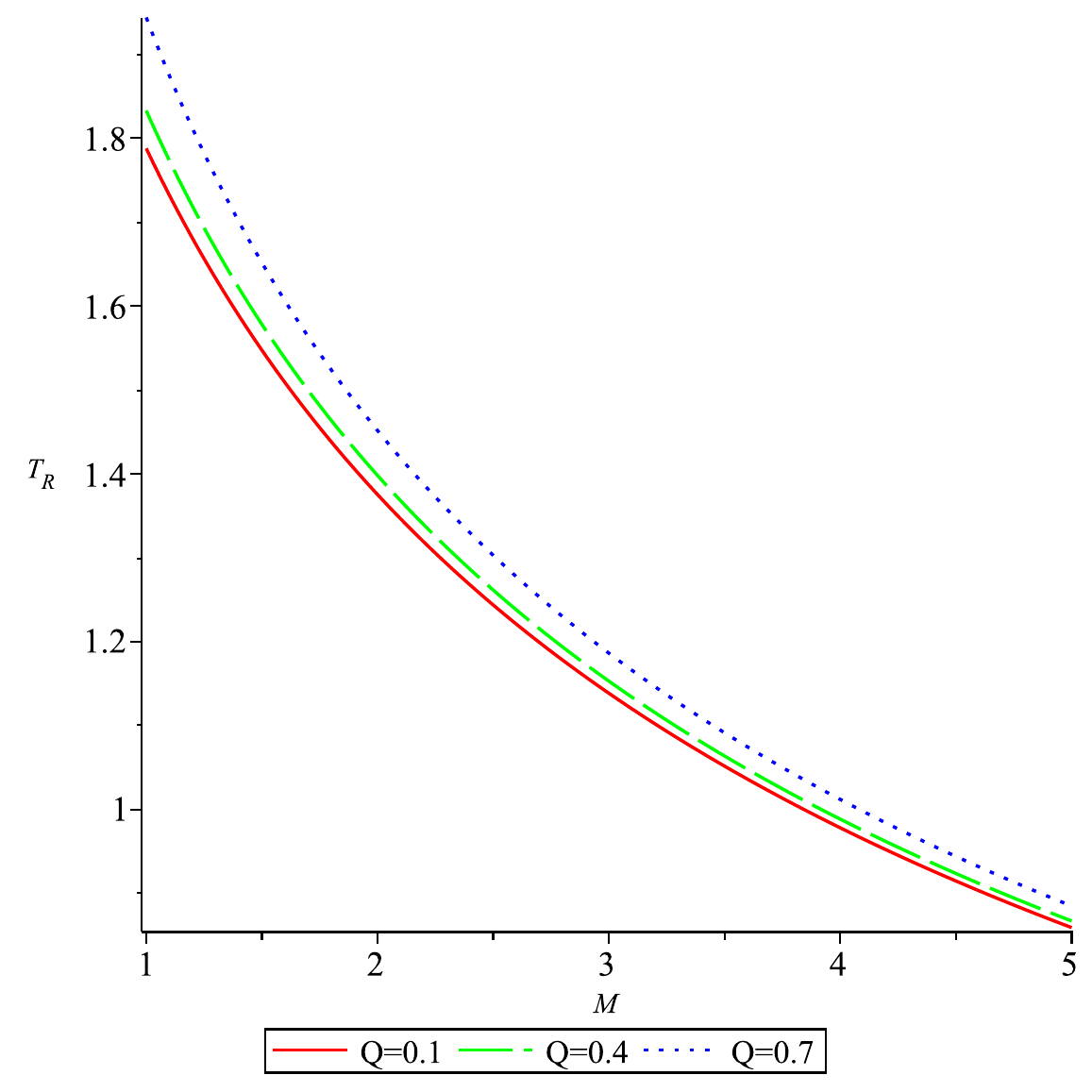}
		\caption{\centering}
		\label{fig:2a}
	\end{subfigure}
	\begin{subfigure}{0.3\linewidth}
		\includegraphics[width=\linewidth]{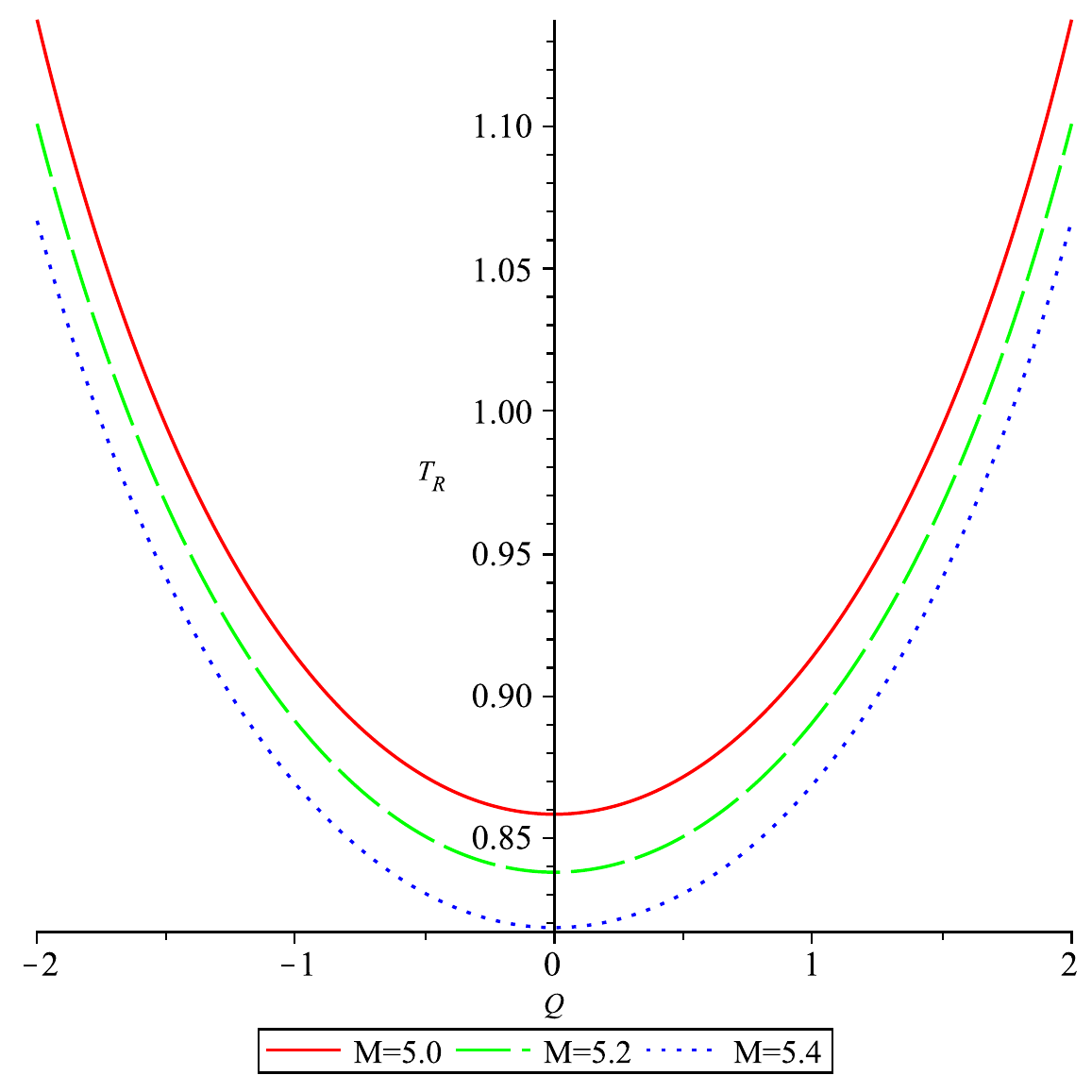}
		\caption{\centering}
		\label{fig:2b}
	\end{subfigure}
	\begin{subfigure}{0.3\linewidth}
		\includegraphics[width=\linewidth]{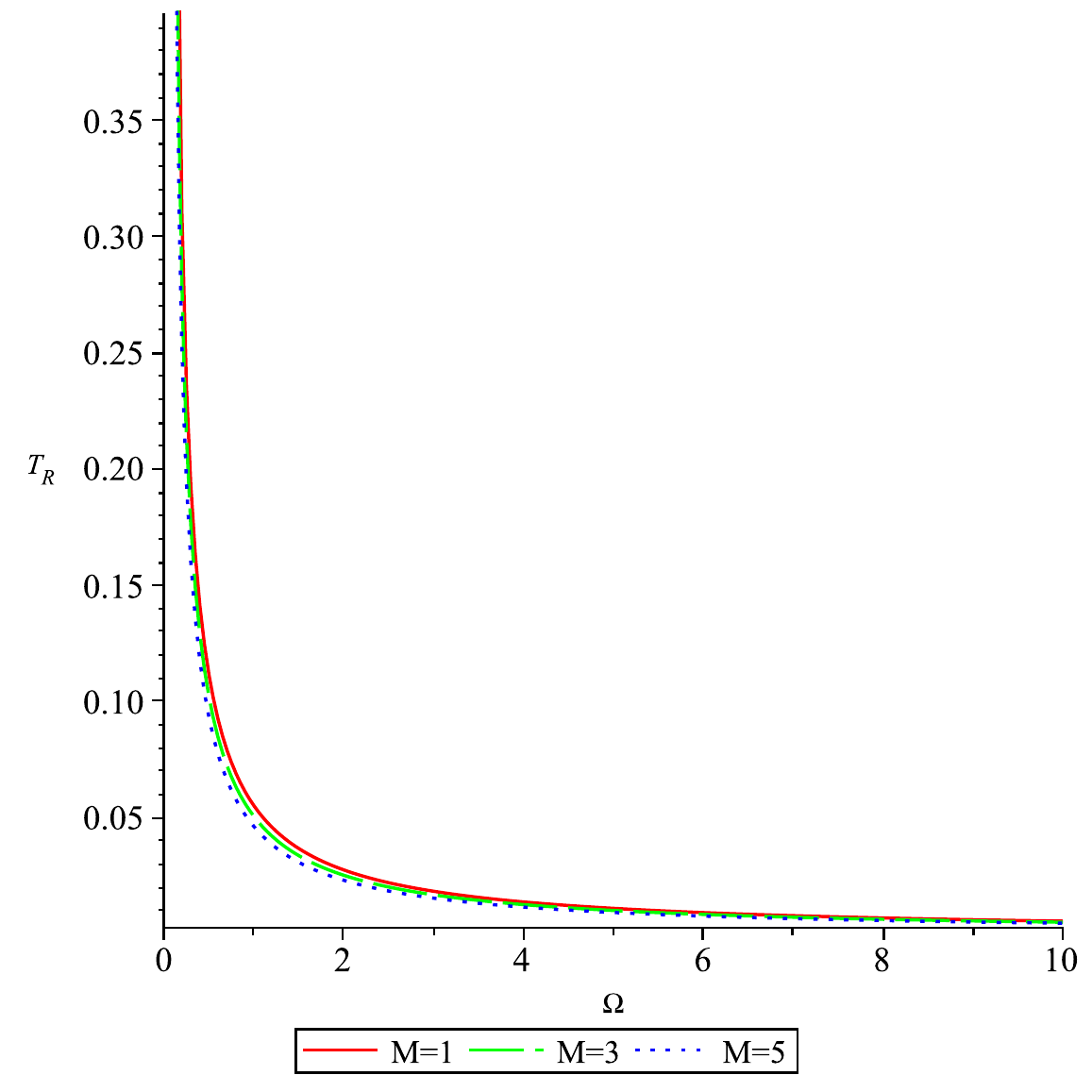}
		\caption{\centering}
		\label{fig:2c}
	\end{subfigure}
	\begin{subfigure}{0.3\linewidth}
		\includegraphics[width=\linewidth]{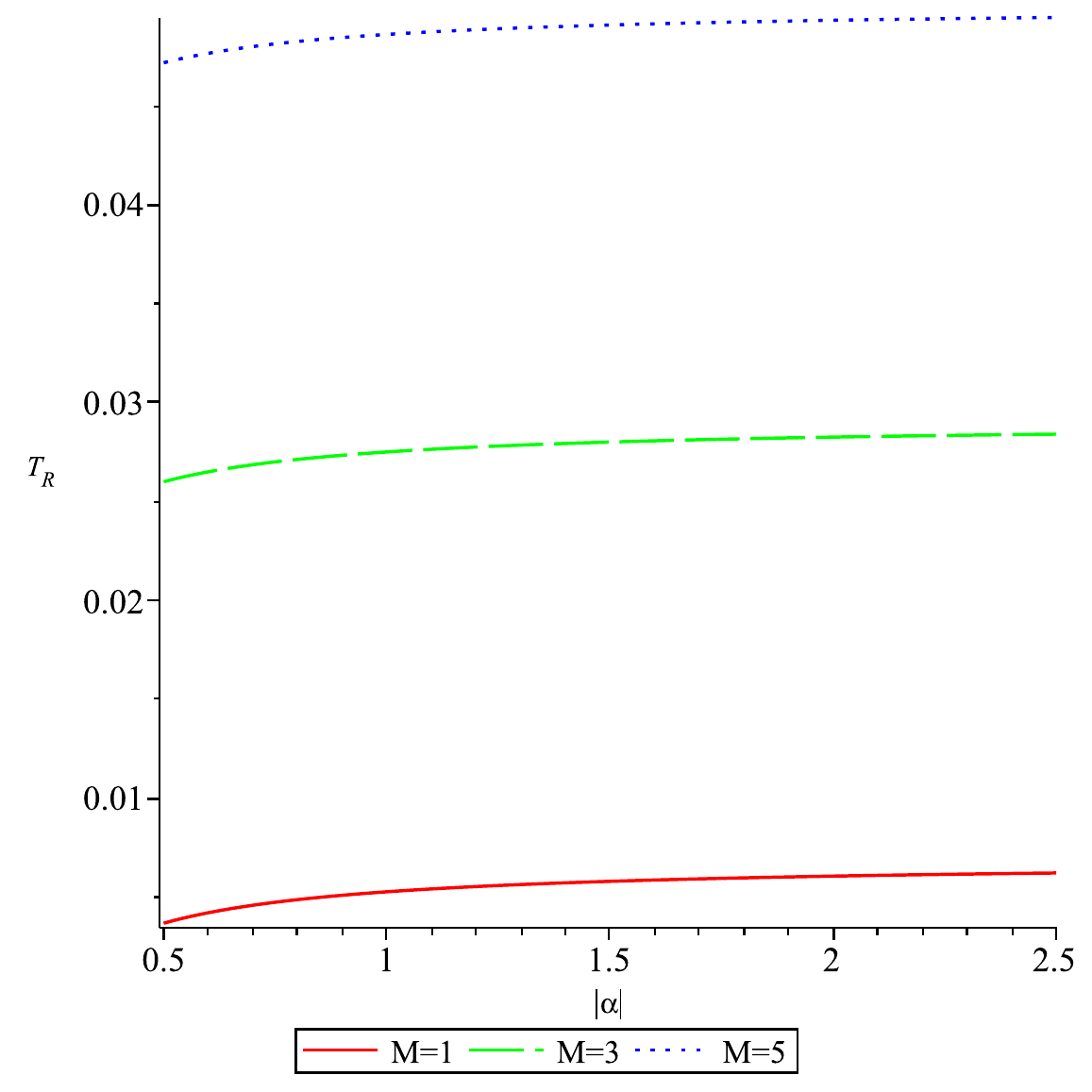}
		\caption{\centering} 
		\label{fig:2d}
	\end{subfigure}
	\begin{subfigure}{0.3\linewidth}
		\includegraphics[width=\linewidth]{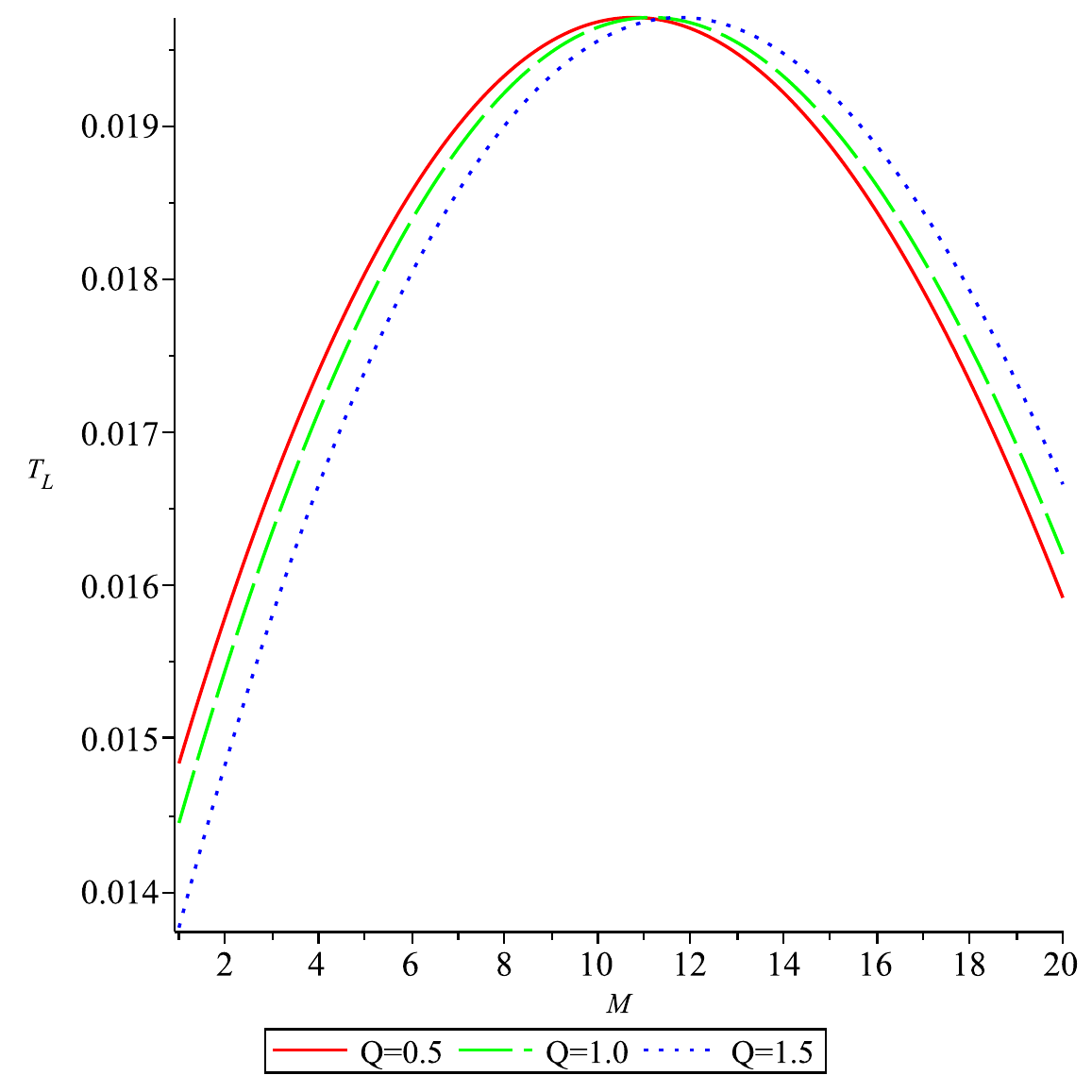}
		\caption{\centering}
		\label{fig:2e}
	\end{subfigure}
	\begin{subfigure}{0.3\linewidth}
		\includegraphics[width=\linewidth]{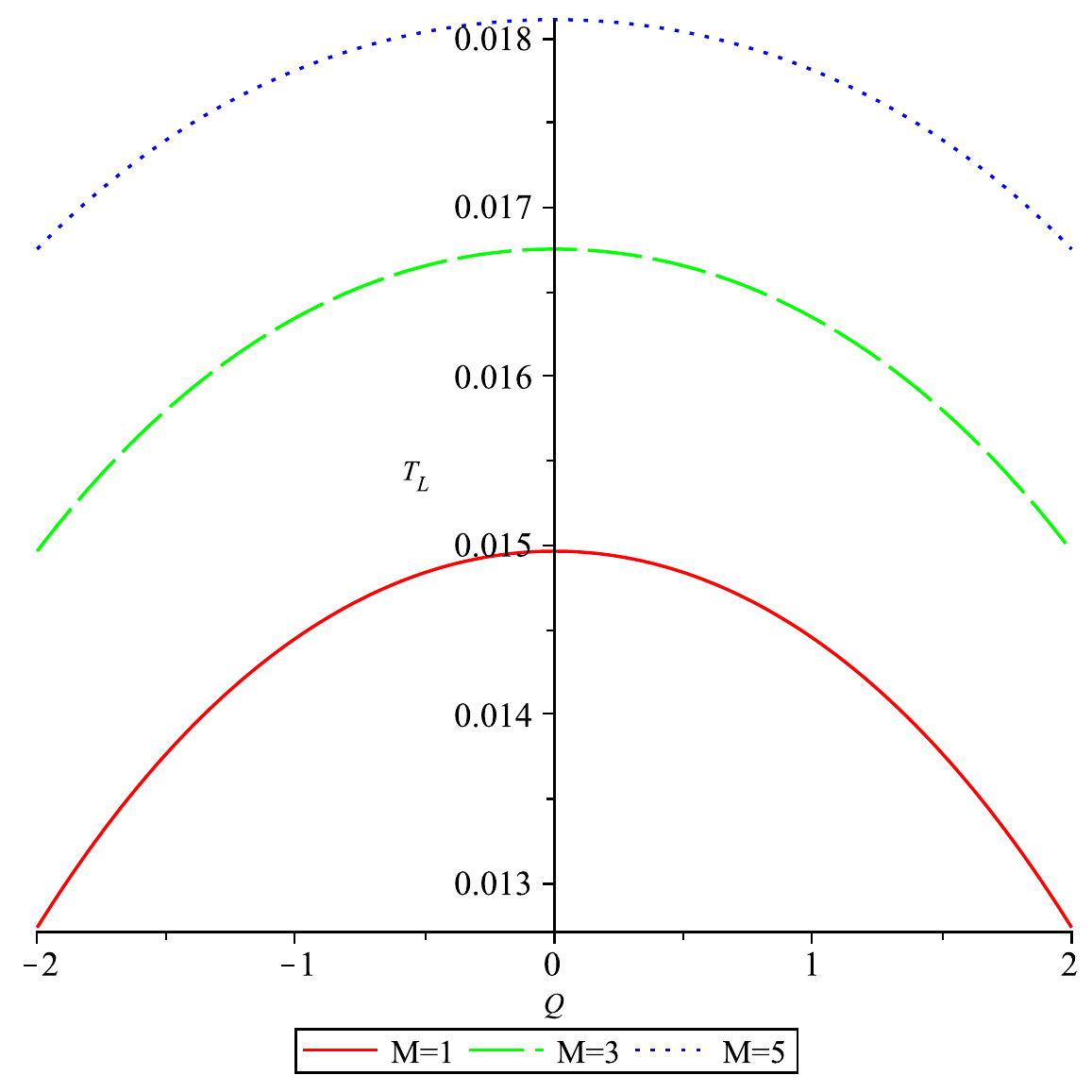}
		\caption{\centering}
		\label{fig:2f}
	\end{subfigure}
	\begin{subfigure}{0.3\linewidth}
		\includegraphics[width=\linewidth]{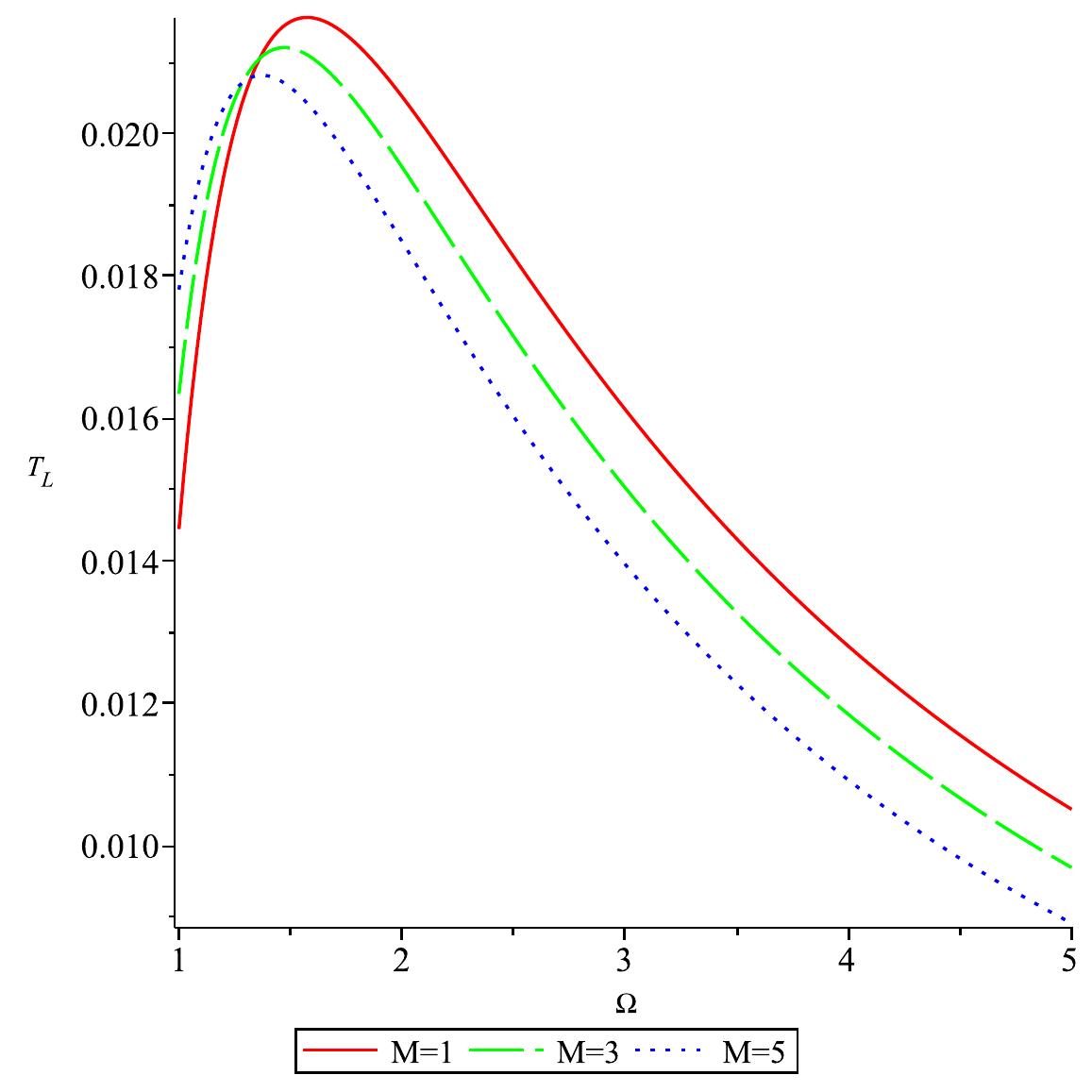}
		\caption{\centering} 
		\label{fig:2g}
	\end{subfigure}
	\begin{subfigure}{0.3\linewidth}
		\includegraphics[width=\linewidth]{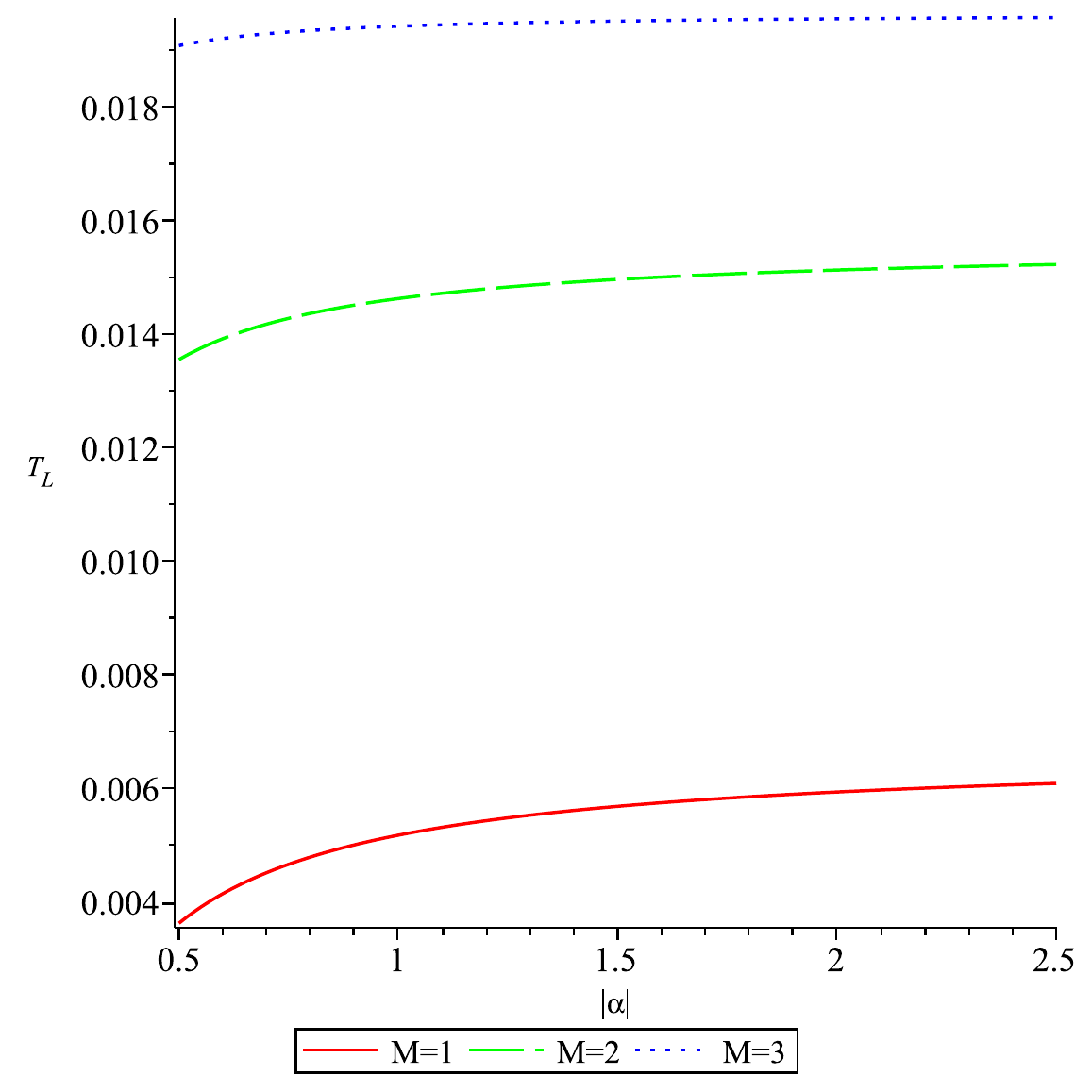}
		\caption{\centering}
		\label{fig:2i}
	\end{subfigure}
	\caption{ The behaviour of $T_R$ and $T_L$ in terms of $M$, $Q$, $\Omega$ and $\vert \alpha\vert $ in $J$ picture, where we set $l=0.1$, $\alpha=0.004$, $r_+=1$ and $r_*=2$. }
	\label{fig:1}
\end{figure}
In Figure (\ref{fig:1}), we show the behaviour of $T_R$ and $T_L$ 
{\textcolor{black}{in the Planck units}}, with respect to the black hole parameters $M$, $Q$, $\Omega$, and $\vert \alpha\vert $ {\textcolor{black}{(all in the Planck units)}} in $J$ picture, {\textcolor{black}{where we consider some arbitrary typical ranges for the black hole parameters and $\alpha$.}}
{\textcolor{black}{We notice different behaviours for the temperatures versus the black hole parameters and the dimensional $f(T)$-parameter $\alpha$. The $T_R$ versus the mass of black hole has a monotonic decreasing behaviour, while $T_L$ increases initially with increasing the mass, reaches to a maximum at $M\simeq 11$ and then decreases with increasing the mass. Moreover,  we notice that $T_R$ and $T_L$ versus the electric charge of black hole, have respectively a local minimum and maximum for charge-less black hole. The $T_R$ has a monotonic decreasing behaviour versus angular velocity of the black hole, while $T_L$ increases initially with increasing the angular velocity, reaches to a maximum at around $\Omega \simeq 1.5$ and then decreases monotonically with increasing the angular velocity. Finally, $T_R$ and $T_L$ both have slightly increasing behaviour with respect to increasing the  $f(T)$-parameter $\alpha$. We should emphasis that the maximum points in some of plots indicate the existence of the upper limit on the temperature of the CFT, while the minimum points refer to the fact that the temperature of the CFT has a lower limit. We also should notice that in the Figure (\ref{fig:1}), we can not take the GR limit, where $\alpha \rightarrow 0$. In fact, the reason is that the non-extremal charged rotating black holes of $f(T)$ gravity do not have any proper limit to approach the black hole solutions of the GR. The only limit that quadratic $f(T)$ black hole can reduce to a GR black hole is setting $\alpha=0$ in (\ref{fTa}), however, the metric functions (\ref{metra}) and (\ref{metrb}) are not well defined in this limit, as $\Lambda_{eff}\rightarrow \infty$.
}}

Second, by considering only the zero-mode of the angular momentum $m=0$ for the scalar probe in the radial equation (\ref{rad2}) and the scalar probe expansion (\ref{scp}), we find the second holographic description which is $Q$ picture. Based on the Casimir operator, the radial equation (\ref{rad2}) can be written as
\begin{equation}
    \mathcal{H}^2R(r)=\Tilde{ \mathcal{H}}^2R(r)=-CR(r),
\end{equation}
and we find the CFT parameters as
\begin{equation}
    n_R^Q=-\frac{S-1}{4UV},
\end{equation}
\begin{equation}
    n_L^Q=-\frac{S+1}{4UV},
\end{equation}
\begin{equation}
    T_R^Q=\frac{\sqrt{6}(Y-1)}{-4\pi XV},
\end{equation}
\begin{equation}
    T_L^Q=\frac{\sqrt{6}(Y+1)}{-4\pi XV},
\end{equation}
where {\textcolor{black}{ the quantities $S$, $U$, $V$, $X$ and $Y$ are given by}}
\begin{equation}
    S=\frac{r_+}{r_*}\sqrt{\frac{r_+^2+2r_+r_*+3r_*^2}{3r_+^2+2r_+r_*+r_*^2}},
\end{equation}
\begin{equation}
    U=\frac{(Q\sqrt{6|\alpha|}+r_+^2)l^2}{r_+^3K(r_+-r_*)(\Xi^2l^2-\Omega^2)}\sqrt{{-K\Omega^2r_+r_*+\Xi^2(3r_+^2+2r_+r_*+r_*^2)}},
\end{equation}
\begin{equation}
    V=-\sqrt{\frac{r_+(K\Omega^2r_*^3-\Xi^2r_+^3-2\Xi^2r_+^2r_*-3\Xi^2r_+r_*^2)}{r_*^2(-3\Xi^2r_+^2+(K\Omega^2-2\Xi^2)r_+r_*-\Xi^2r_*^2))}}+\frac{r_+}{r_*}\sqrt{\frac{r_+^2+2r_+r_*+3r_*^2}{3r_+^2+2r_+r_*+r_*^2}}, \label{V}
\end{equation}
\begin{equation}
    X=\frac{(Q\sqrt{6|\alpha|}+r_+^2)Q(\sqrt{6}r_+^2+2Q\sqrt{|\alpha|})}{(r_+-r_*)r_+^{6}K}\sqrt{
    {
    (3r_+^2+2r_+r_*+r_*^2)
    }
    }, \label{X}
\end{equation}
\begin{equation}
    Y=\sqrt{\frac{r_+(K\Omega^2r_*^3-\Xi^2r_+^3-2\Xi^2r_+^2r_*-3\Xi^2r_+r_*^2)}{r_*^2(-3\Xi^2r_+^2+(K\Omega^2-2\Xi^2)r_+r_*-\Xi^2r_*^2))}}. \label{Y}
\end{equation}
\begin{figure}[h]
	\centering
	\begin{subfigure}{0.3\linewidth}
		\includegraphics[width=\linewidth]{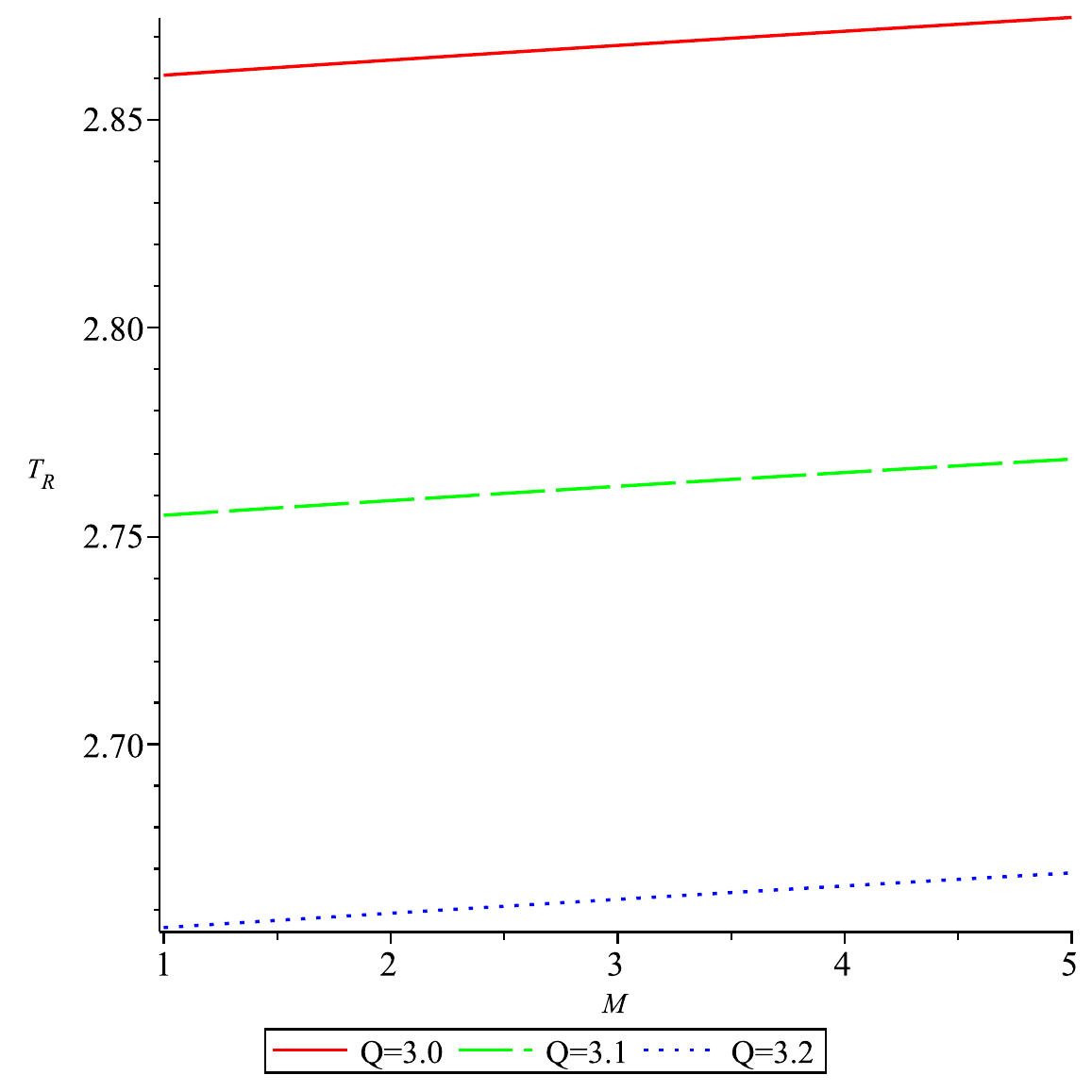}
		\caption{\centering}
		\label{fig:2a}
	\end{subfigure}
	\begin{subfigure}{0.3\linewidth}
		\includegraphics[width=\linewidth]{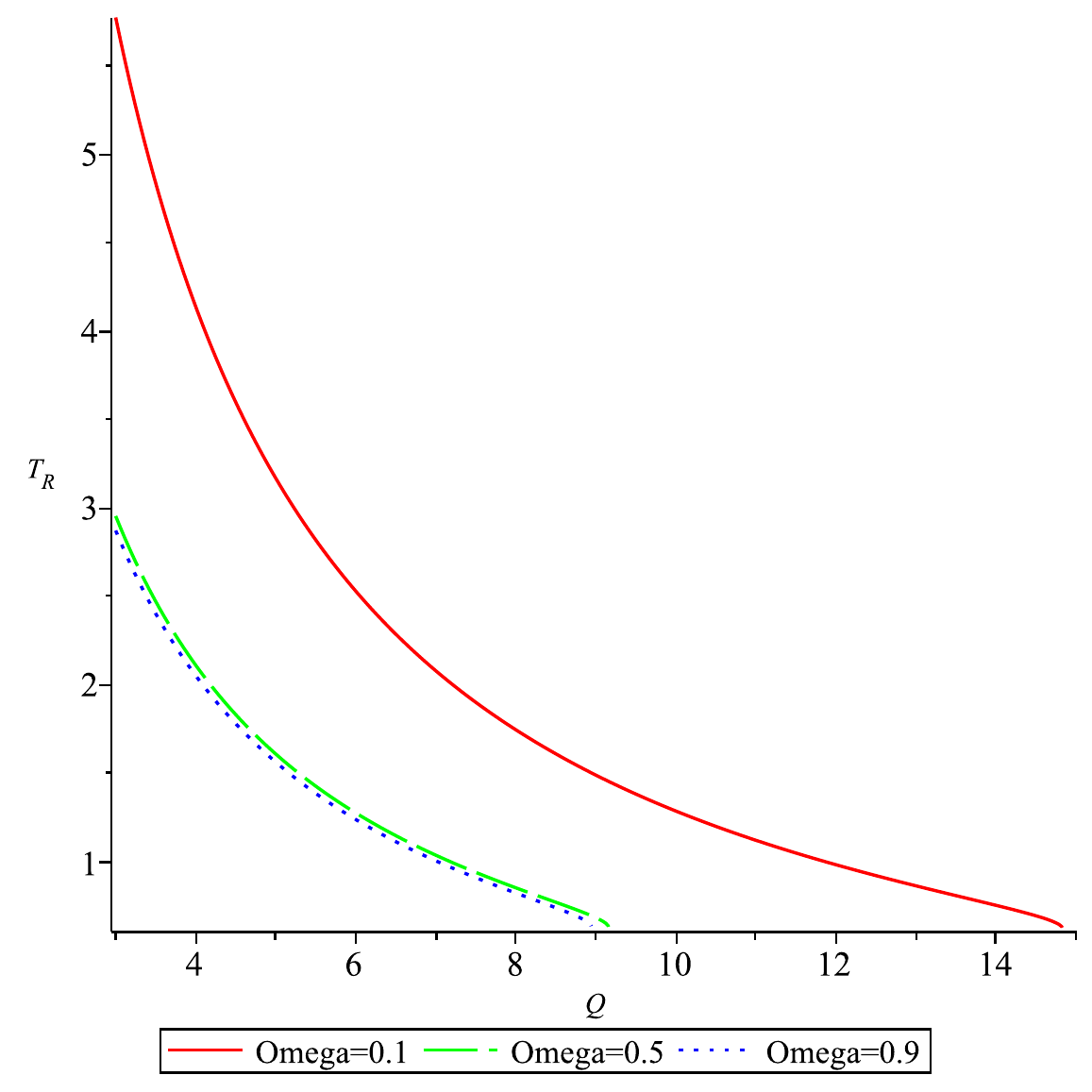}
		\caption{\centering}
		\label{fig:2b}
	\end{subfigure}
	\begin{subfigure}{0.3\linewidth}
		\includegraphics[width=\linewidth]{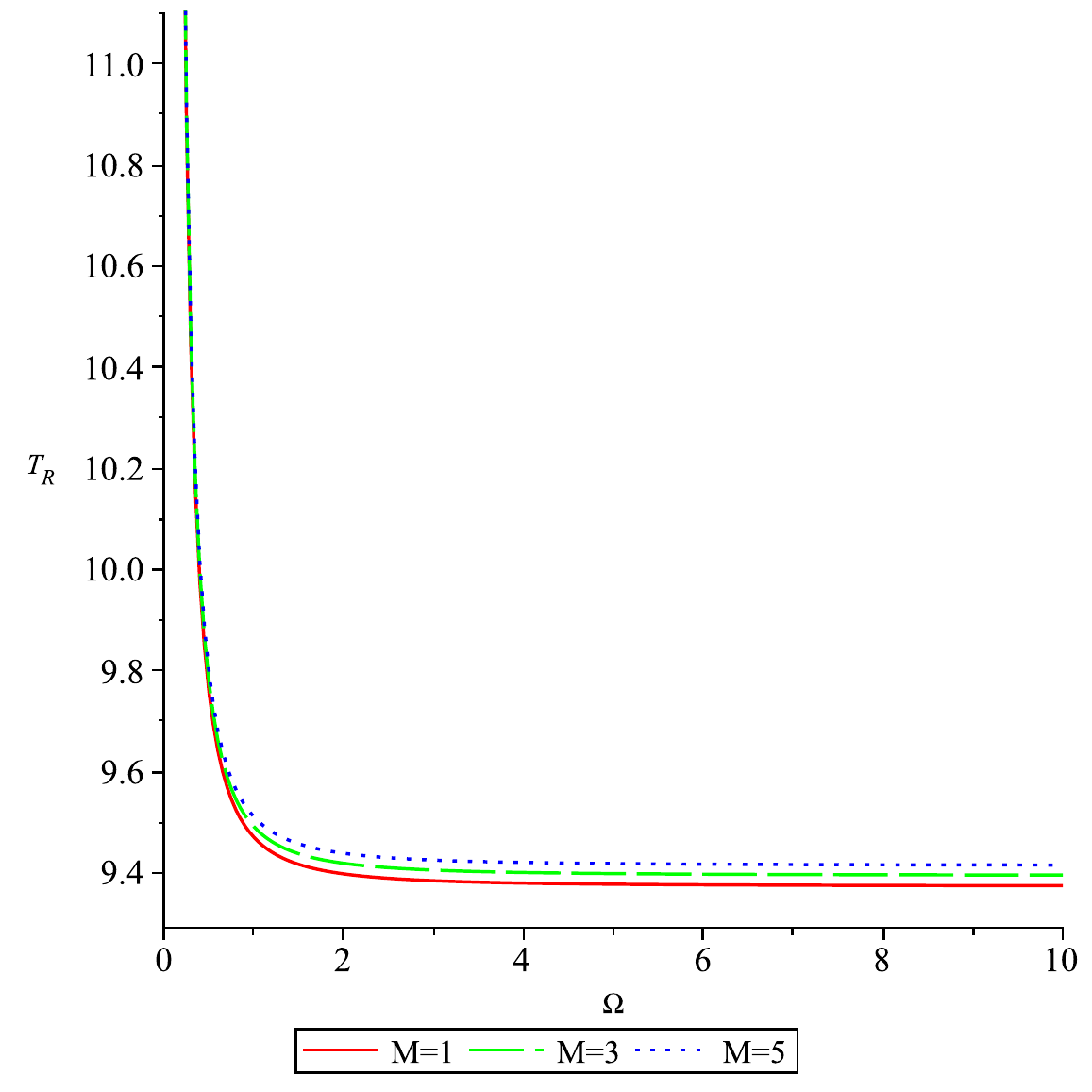}
		\caption{\centering}
		\label{fig:2c}
	\end{subfigure}
	\begin{subfigure}{0.3\linewidth}
		\includegraphics[width=\linewidth]{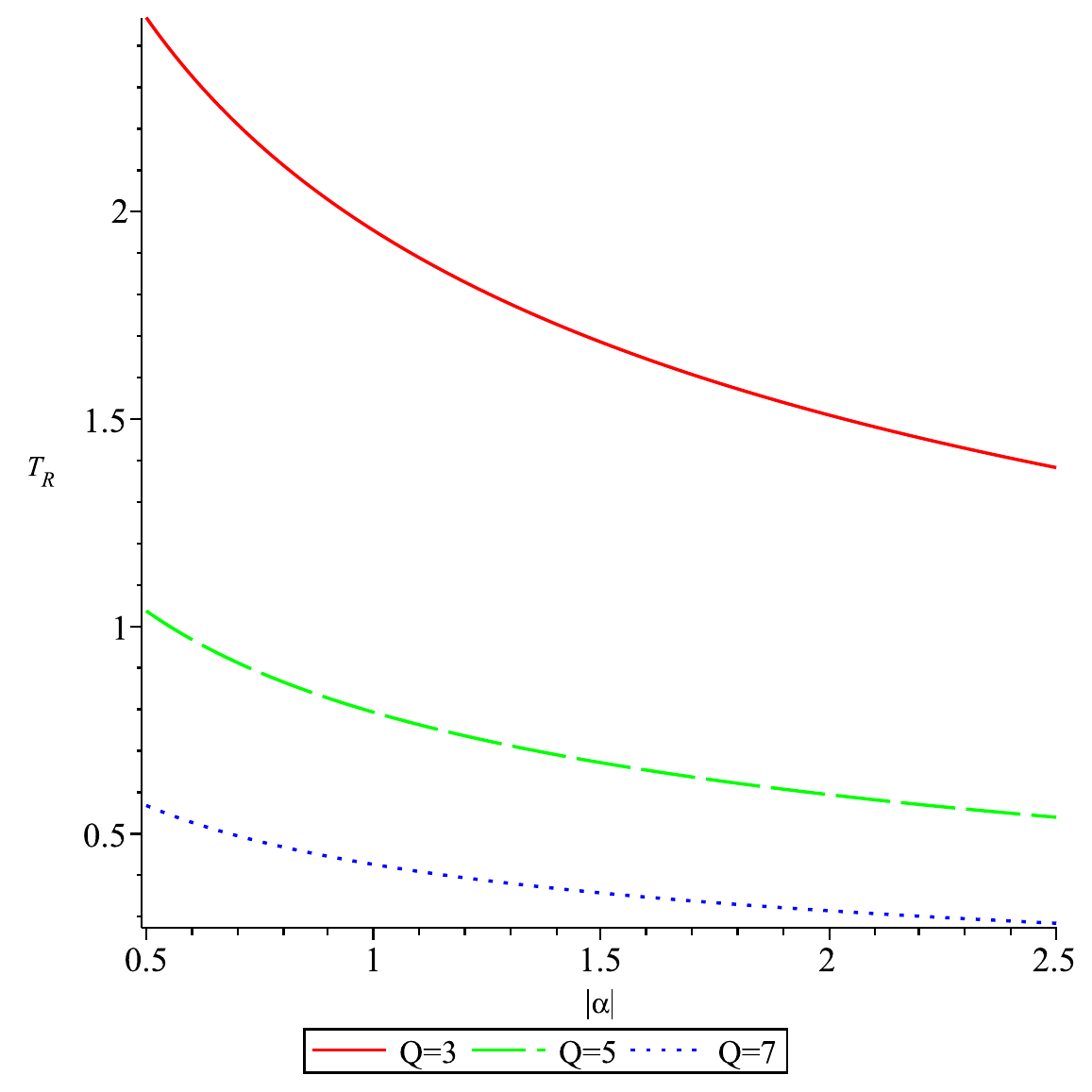}
		\caption{\centering} 
		\label{fig:2d}
	\end{subfigure}
	\begin{subfigure}{0.3\linewidth}
		\includegraphics[width=\linewidth]{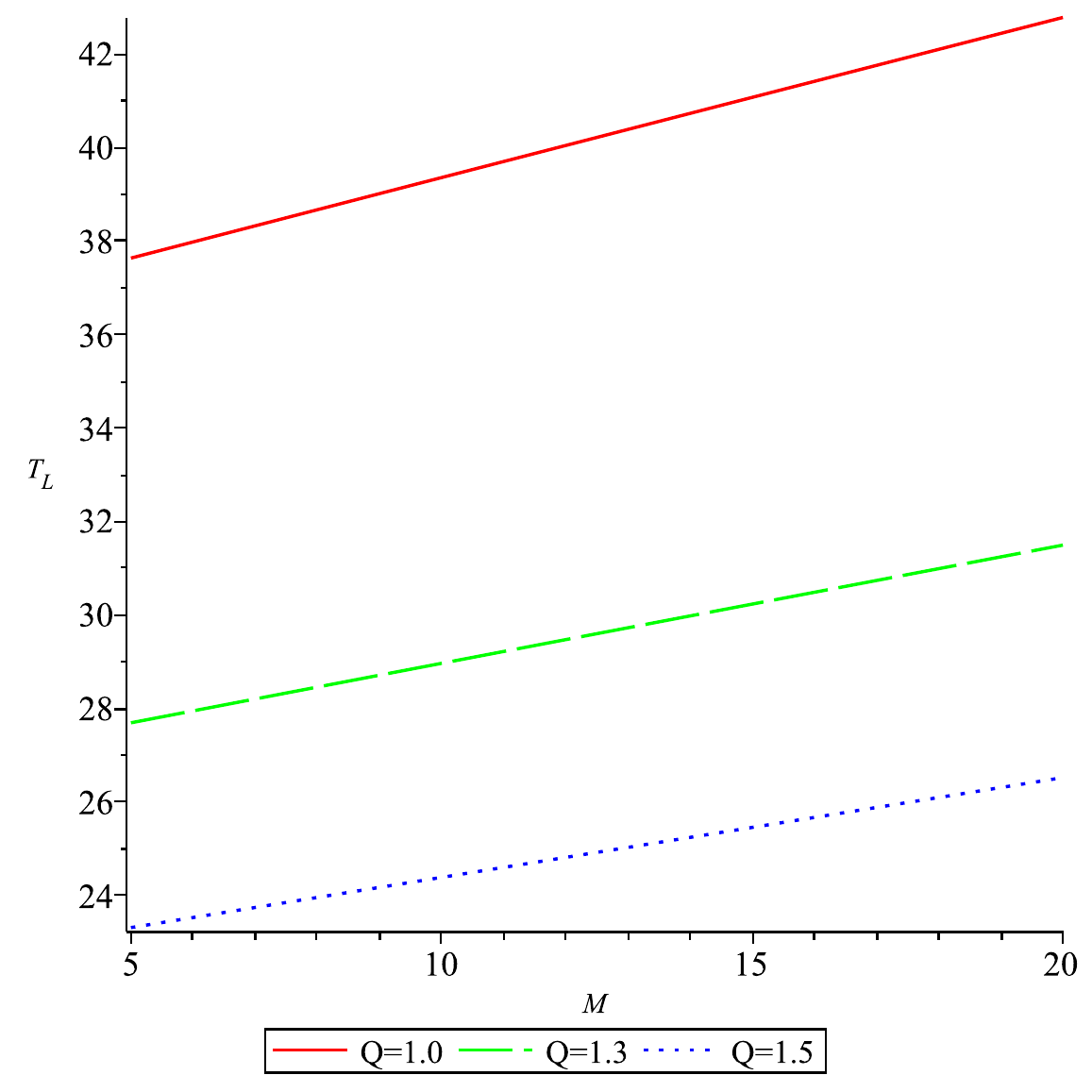}
		\caption{\centering}
		\label{fig:2e}
	\end{subfigure}
	\begin{subfigure}{0.3\linewidth}
		\includegraphics[width=\linewidth]{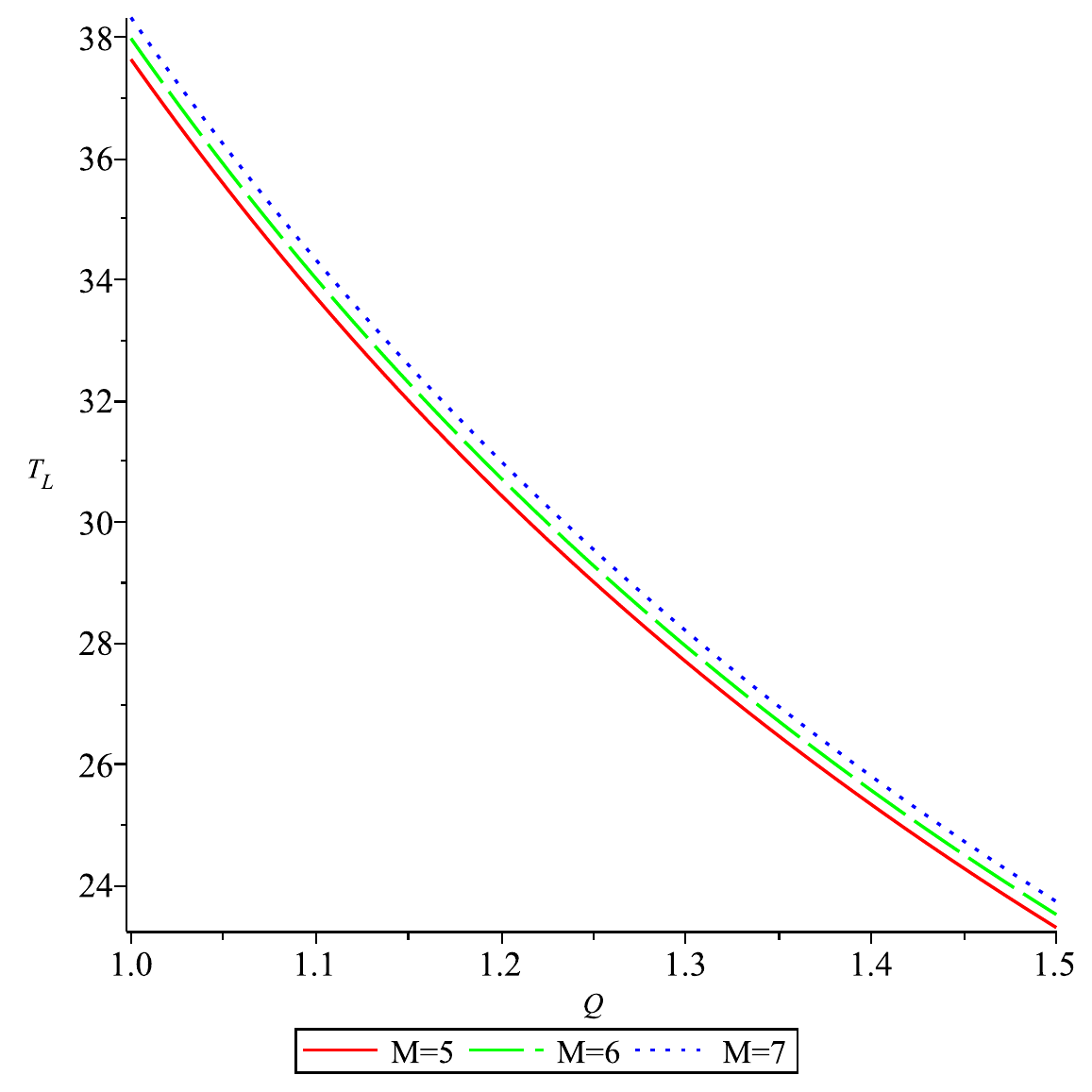}
		\caption{\centering}
		\label{fig:2f}
	\end{subfigure}
	\begin{subfigure}{0.3\linewidth}
		\includegraphics[width=\linewidth]{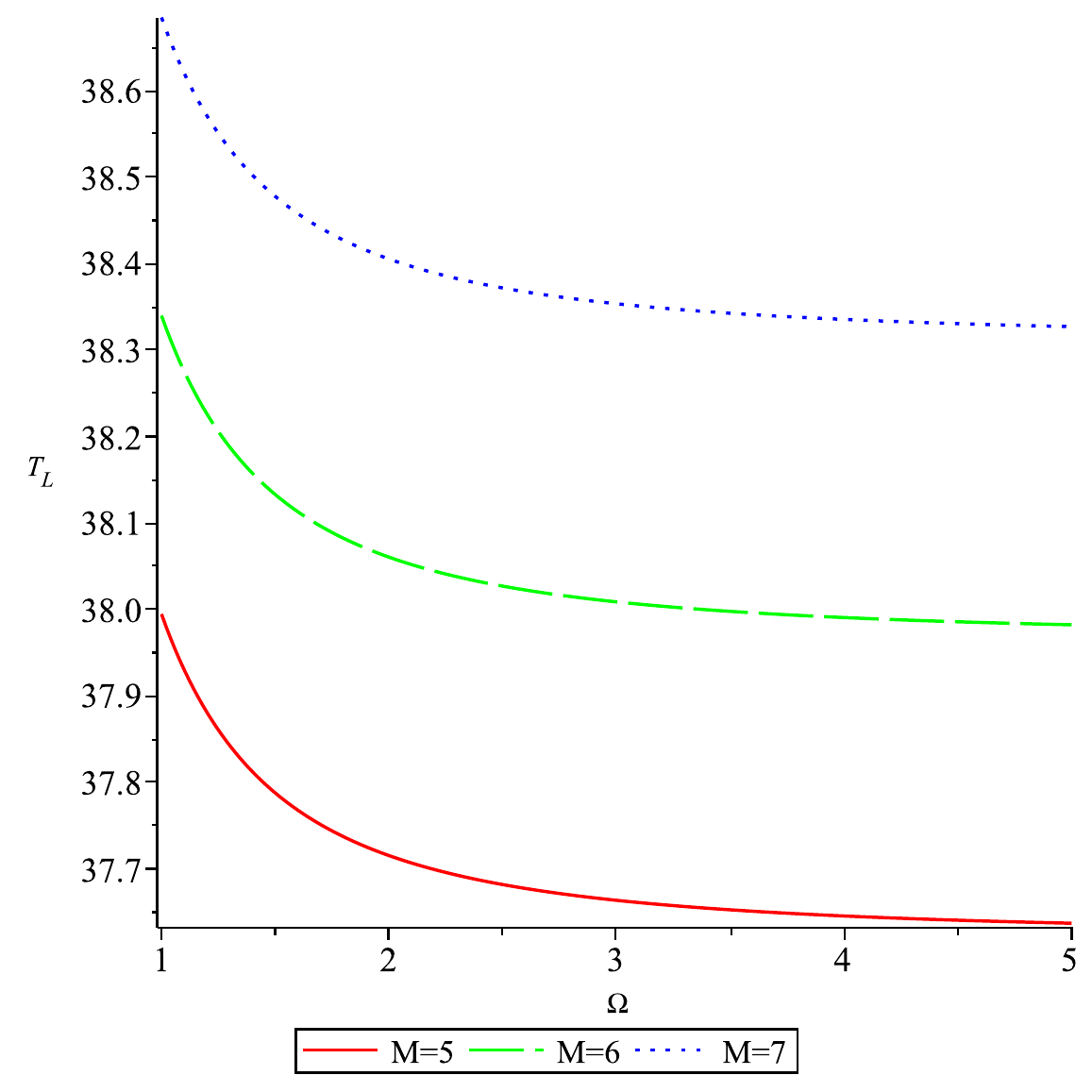}
		\caption{\centering} 
		\label{fig:2g}
	\end{subfigure}
	\begin{subfigure}{0.3\linewidth}
		\includegraphics[width=\linewidth]{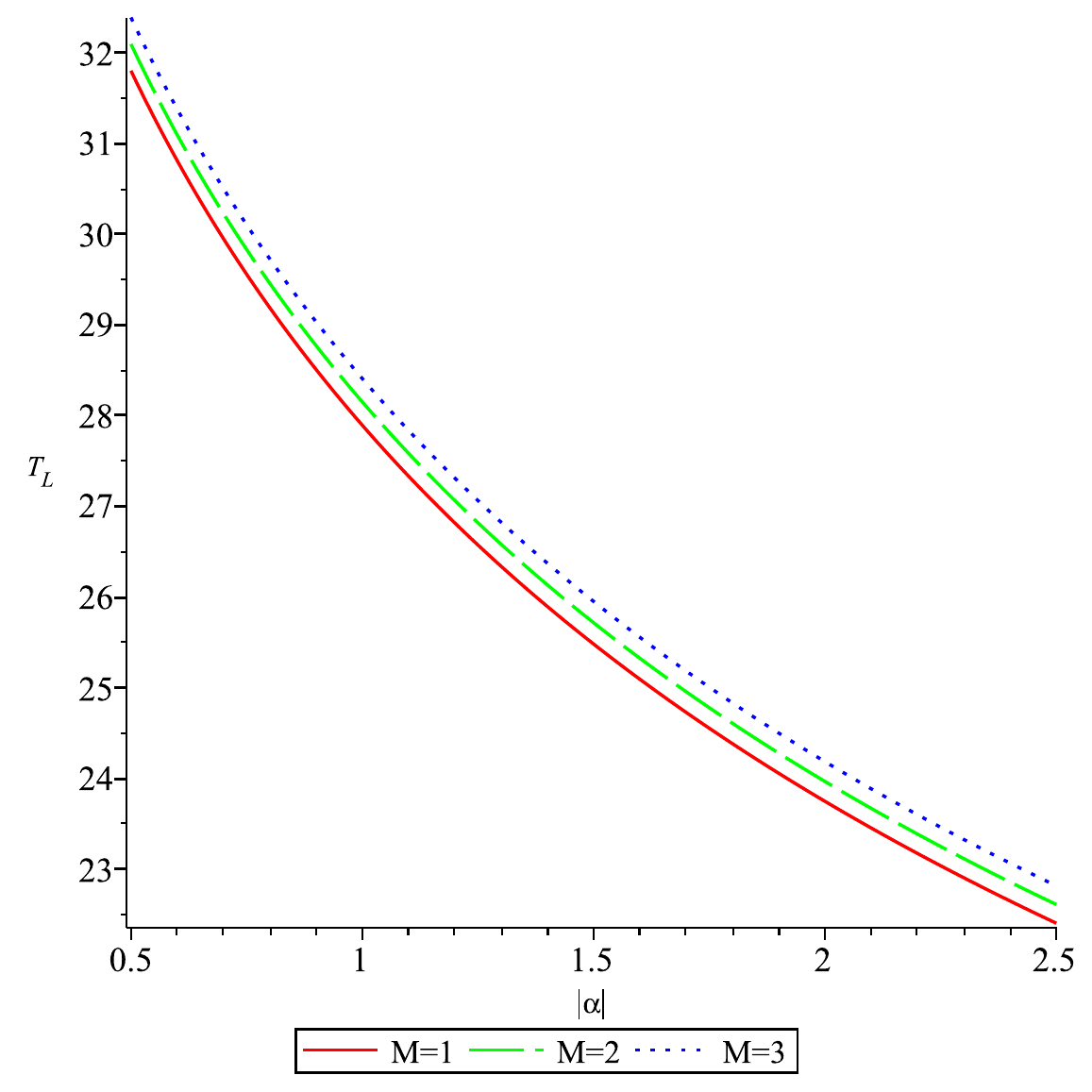}
		\caption{\centering}
		\label{fig:2i}
	\end{subfigure}
	\caption{ The behaviour of $T_R$ and $T_L$ in terms of $M$, $Q$, $\Omega$ and $\vert \alpha\vert $ in $Q$ picture, where we set $l=0.1$, $\alpha=0.004$, $r_+=1$ and $r_*=2$. }
	\label{fig:Q}
\end{figure}
In Figure (\ref{fig:Q}), we show the behaviour of $T_R$ and $T_L$ {\textcolor{black}{in the Planck units}}, in terms of the black hole parameters $M$, $Q$, $\Omega$, and $\vert\alpha\vert$ {\textcolor{black}{(all in the Planck units)}} in $Q$ picture, {\textcolor{black}{where we consider some arbitrary typical ranges for the black hole parameters and $\alpha$.}}
{\textcolor{black}{We notice different behaviours for the temperatures versus the black hole parameters and the dimensional $f(T)$-parameter $\alpha$. The $T_R$ versus the mass of black hole has a slightly monotonic increasing behaviour, while $T_L$ has a more monotonic increasing behaviour versus the mass.
Moreover,  we notice that $T_R$ and $T_L$ versus the electric charge of black hole, have both monotonic decreasing behaviour with respect to the electric charge of the black hole. The $T_R$ has a sharp decreasing behaviour versus angular velocity of the black hole, while $T_L$ has a moderate decreasing behaviour with increasing the angular velocity of the black hole. 
Both  $T_R$ and $T_L$  have decreasing behaviour with respect to increasing the  $f(T)$-parameter $\alpha$. We explicitly notice the different behaviours for the $T_R$ and $T_L$ versus the black hole parameters and  the dimensional $f(T)$-parameter $\alpha$ in $J$ and $Q$ pictures.
}}

Combining the two $U(1)$ symmetries in $J$ and $Q$ picture, we find the corresponding CFT quantities in the general picture. The modular group is given by \cite{OLD11}
\begin{equation}
    \binom{\phi'}{\chi'}=\binom{\gamma \ \lambda}{\eta \ \tau}\binom{\phi}{\chi}. \label{e74}
\end{equation}
Applying this transformation, we find
\begin{equation}
    m=\gamma m'+\eta q',
\end{equation}
\begin{equation}
    q=\lambda m'+\tau q',
\end{equation}
where $\gamma$, $\lambda$, $\eta$ and $\tau$ parameterize the $SL(2,\mathbb{Z})$ group elements. The radial equation becomes
\begin{equation}
    \frac{d}{dr}\{\left( {r - {r_ + }} \right)\left( {r - {r_ * }} \right)\frac{d}{dr}R\left( r \right)\}+ \left[ {\left( {\frac{{{r_ + } - {r_ * }}}{{r - {r_ + }}}} \right)\mathcal{A} + \left( {\frac{{{r_ + } - {r_ * }}}{{r - {r_ * }}}} \right)\mathcal{B} + \mathcal{C}} \right]R\left( r \right) = 0, \label{radextj}
\end{equation}
where $\mathcal{A}$, $\mathcal{B}$ and $\mathcal{C}$ can be found by performing the transformation (\ref{e74}) on equations (\ref{A1}), (\ref{B1}) and (\ref{C1}), respectively.

Moreover, we can define the $J'$ and $Q'$ descriptions in the general picture, by considering $q'=0$ and $m'=0$, respectively. 
By comparing the radial equation (\ref{radextj}) and the Casimir operator (\ref{csm}), we find the following CFT parameters in $J'$ picture

\begin{equation}
    n_R^{J'}=\frac{1-a_1}{a_1-b_1}(-4\mathbf{B})^{-1/2}, \label{e78}
\end{equation}
\begin{equation}
    n_L^{J'}=\frac{a_1+1}{a_1-1}n_R^{J'},
\end{equation}
\begin{equation}
    T_R^{J'}=\frac{ n_L^{J'}-n_R^{J'}}{( n_L^{J'}-(\frac{b_1+1}{b_1-1})n_R^{J'})4\pi}(-\mathbf{C})^{-1/2},
\end{equation}
\begin{equation}
    T_L^{J'}=(\frac{b_1+1}{b_1-1})T_R^{J'}, \label{e81}
\end{equation}
where {\textcolor{black}{the quantities $a_1$, $b_1$, $\mathbf{A}$, $\mathbf{B}$, $\mathbf{C}$ and $\mathbf{D}$ are given by}}
\begin{equation}
    a_1=(-\frac{\mathbf{D}}{\mathbf{C}})^{1/2},
\end{equation}
\begin{equation}
    b_1=(-\frac{\mathbf{A}}{\mathbf{B}})^{1/2},
\end{equation}
\begin{equation}
    \mathbf{A}=\frac{(Q\sqrt{6|\alpha|}+r_+^2)^2(-3Kr_*^3l^4\Omega^2r_+^5+3\Xi^2l^4r_+^8+6\Xi^2l^4r_*r_+^7+9\Xi^2l^4r_*^2r_+^6)}{3r_+^8(-\Xi l+\Omega)^2(\Xi l+\Omega)^2(r_+-r_*)^2K^2 r_*^2},
\end{equation}
\begin{equation}
    \mathbf{B}=\frac{(Q\sqrt{6|\alpha|}+r_+^2)^2(-9\Xi^2l^4r_+^8+3(K\Omega^2-2\Xi^2)l^4r_*r_+^7-3\Xi^2l^4r_*^2r_+^6)}{3K^2r_+^{10}(r_+-r_*)^2(-\Xi l+\Omega)^2(\Xi l+\Omega)^2},
\end{equation}
\begin{eqnarray}
    \mathbf{C}&=& \frac{(Q\sqrt{6|\alpha|}+r_+^2)^2}{3K^2 r_+^{10}(r_+-r_*)^2(\Xi^2 l^2-\Omega^2)^2}(6\sqrt{6|\alpha|}Q^2\lambda(\Omega\gamma r_++\lambda Q(-\Xi^2 l^2+\Omega^2)) \nonumber\\
    &\times& (\Xi^2 l^2-\Omega^2)r_+^2(r_+^2+2/3r_+r_*+1/3r_*^2)-9\Omega^2\gamma^2r_+^8+3((6\Omega\Xi^2l^2-6\Omega^3)Q\lambda \nonumber\\
&+&\gamma r_*(K\Xi^2l^4-2\Omega^2))\gamma r_+^7-3((-3\Xi^2 l^2+3\Omega^2)Q\lambda+\Omega\gamma r_*)(\lambda Q(-\Xi^2 l^2+\Omega^2)+\Omega\gamma r_*)r_+^6 \nonumber\\
&+& 6Q\lambda r_*(\lambda Q(-\Xi^2 l^2+\Omega^2)+\Omega\gamma r_*)(\Xi^2 l^2-\Omega^2)r_+^5-3\lambda^2Q^2r_*^2(\Xi^2 l^2-\Omega^2)^2r_+^4 \nonumber\\
&-& 6\lambda^2Q^4\alpha(\Xi^2 l^2-\Omega^2)^2r_+^2-4\lambda^2Q^4r_*\alpha(\Xi^2 l^2-\Omega^2)^2r_+-2\lambda^2Q^4r_*^2\alpha(\Xi^2 l^2-\Omega^2)^2),
\end{eqnarray}
\begin{eqnarray}
    \mathbf{D}&=& \frac{(Q\sqrt{6|\alpha|}+r_+^2)^2}{3r_+^8(\Xi^2l^2-\Omega^2)^2(r_+-r_*)^2K^2 r_*^2}(-2(r_+^2+2r_+r_*+3r_*^2)(\Xi^2 l^2-\Omega^2)r_+^2Q^2\lambda\sqrt{6|\alpha|} \nonumber\\
    &\times& (\Omega\gamma r_++\lambda Q(-\Xi^2l^2+\Omega^2))+3\Omega^2\gamma^2r_+^8+6(\lambda Q(-\Xi^2l^2+\Omega^2)+\Omega\gamma r_*)\Omega\gamma r_+^7+9(\lambda Q \nonumber\\
    &\times& (-\Xi^2l^2+\Omega^2)+\Omega\gamma r_*)(Q(-1/3\Xi^2l^2+1/3\Omega^2)\lambda+\Omega\gamma r_*)r_+^6-3r_*(-2Q(\Xi^2 l^2-\Omega^2)^2\lambda^2 \nonumber\\
    &+& (6\Omega\Xi^2\gamma l^2r_*-6\Omega^3\gamma r_*)Q\lambda+K\gamma^2l^4\Xi^2r_*^2)r_+^5+9\lambda^2Q^2r_*^2(\Xi^2 l^2-\Omega^2)^2r_+^4 \nonumber\\
    &+& 2\lambda^2Q^4\alpha(\Xi^2 l^2-\Omega^2)^2r_+^2+4\lambda^2Q^4r_*\alpha(\Xi^2 l^2-\Omega^2)^2r_++6\lambda^2Q^4r_*^2\alpha(\Xi^2 l^2-\Omega^2)^2).
\end{eqnarray}
We note that the CFT parameters (\ref{e78})-(\ref{e81}) reduce exactly to the corresponding results in $J$ picture, where  we consider the unit element of $SL(2,\mathbb{Z})$  with $\gamma=1$ and $\lambda=0$. 

Moreover, by setting $m'=0$, the radial equation becomes similar to (\ref{radextj}). By the replacement of $(\gamma,\lambda,m')$ to $(\eta,\tau,q')$ in the radial equation (\ref{radextj}) as well as the results (\ref{e78})-(\ref{e81}), we find the CFT parameters in $Q'$ picture.  We have explicitly checked out that the CFT parameters in $Q'$ picture, reduce exactly to the corresponding results in $Q$ picture, where  we consider the unit element of $SL(2,\mathbb{Z})$  with $\eta=0$ and $\tau=1$. 

We note that $SL(2,\mathbb{R})_L \times SL(2,\mathbb{R})_R$ algebra is only a local hidden conformal symmetry, which is generated by the vector fields (\ref{opi})-(\ref{opf}) and is not globally defined. This symmetry is broken spontaneously by the periodic identification of the angular coordinate $\phi$ \cite{OLD134},
\begin{equation}
    \phi \sim \phi+2\pi .
\end{equation}
Therefore, we may assume that the near horizon dynamics can be described by a dual CFT correspondence. For this purpose, we find the left and right charges of the  dual CFT, from which we can calculate and compare the microscopic entropy description of the dual CFT and the Bekenstein-Hawking entropy of the black hole. According to the Cardy formula, the entropy of the CFT  is given by
\begin{equation}
    S_{CFT}=\frac{\pi^2}{3}(c_LT_L+c_RT_R), \label{cardy}
\end{equation}
where $c_L$ and $c_R$ are the  left and right central charges. We show that the microscopic entropy of the dual CFT is in agreement with the Bekenstein-Hawking entropy for the four-dimensional charged rotating  AdS black hole in {\textcolor{black}{quadratic}} $f(T)$-Maxwell theory. 

\begin{figure}[H]
	\centering
		\begin{subfigure}{0.3\linewidth}
		\includegraphics[width=\linewidth]{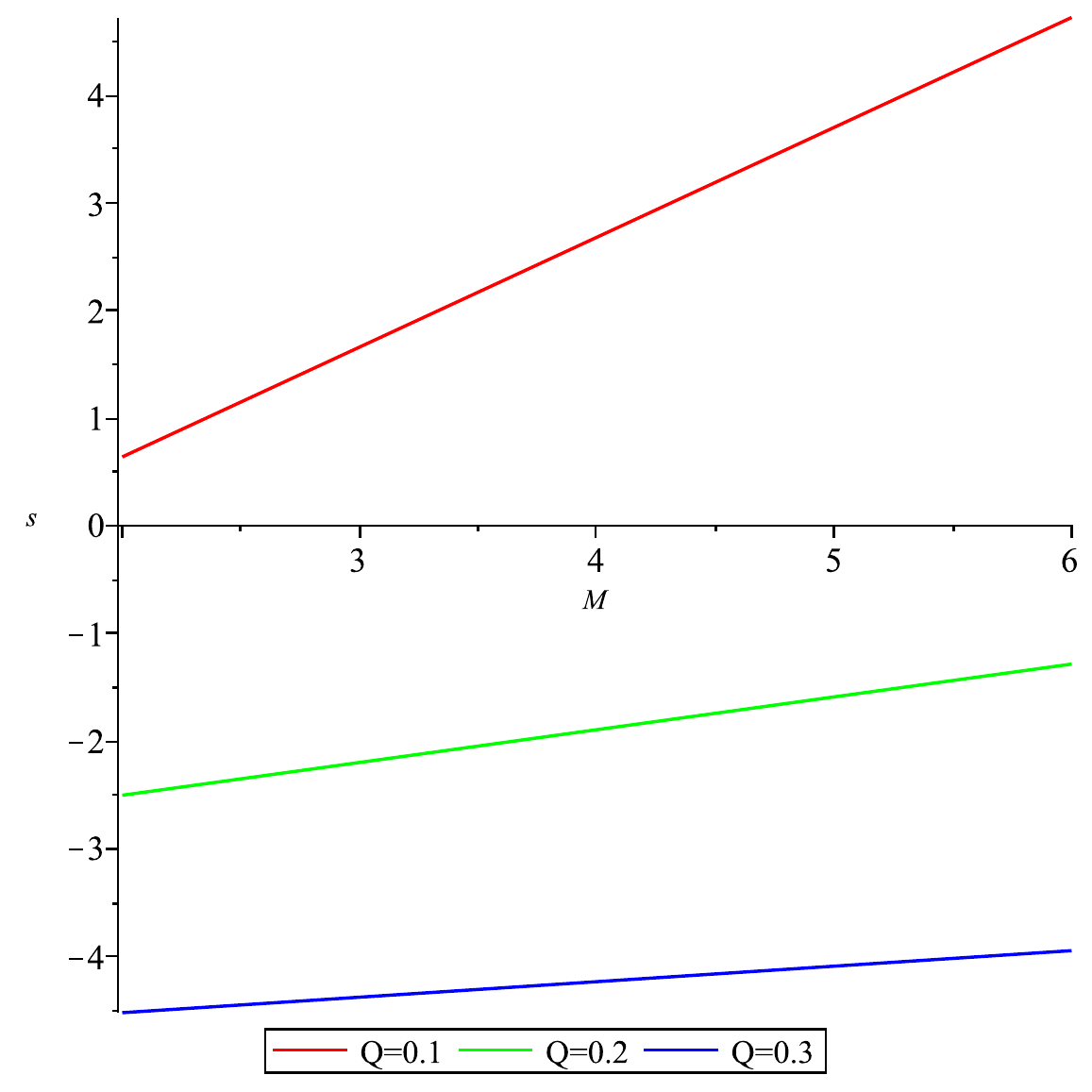}
		\caption{\centering}
		\label{fig:2c}
	\end{subfigure}
	\begin{subfigure}{0.3\linewidth}
		\includegraphics[width=\linewidth]{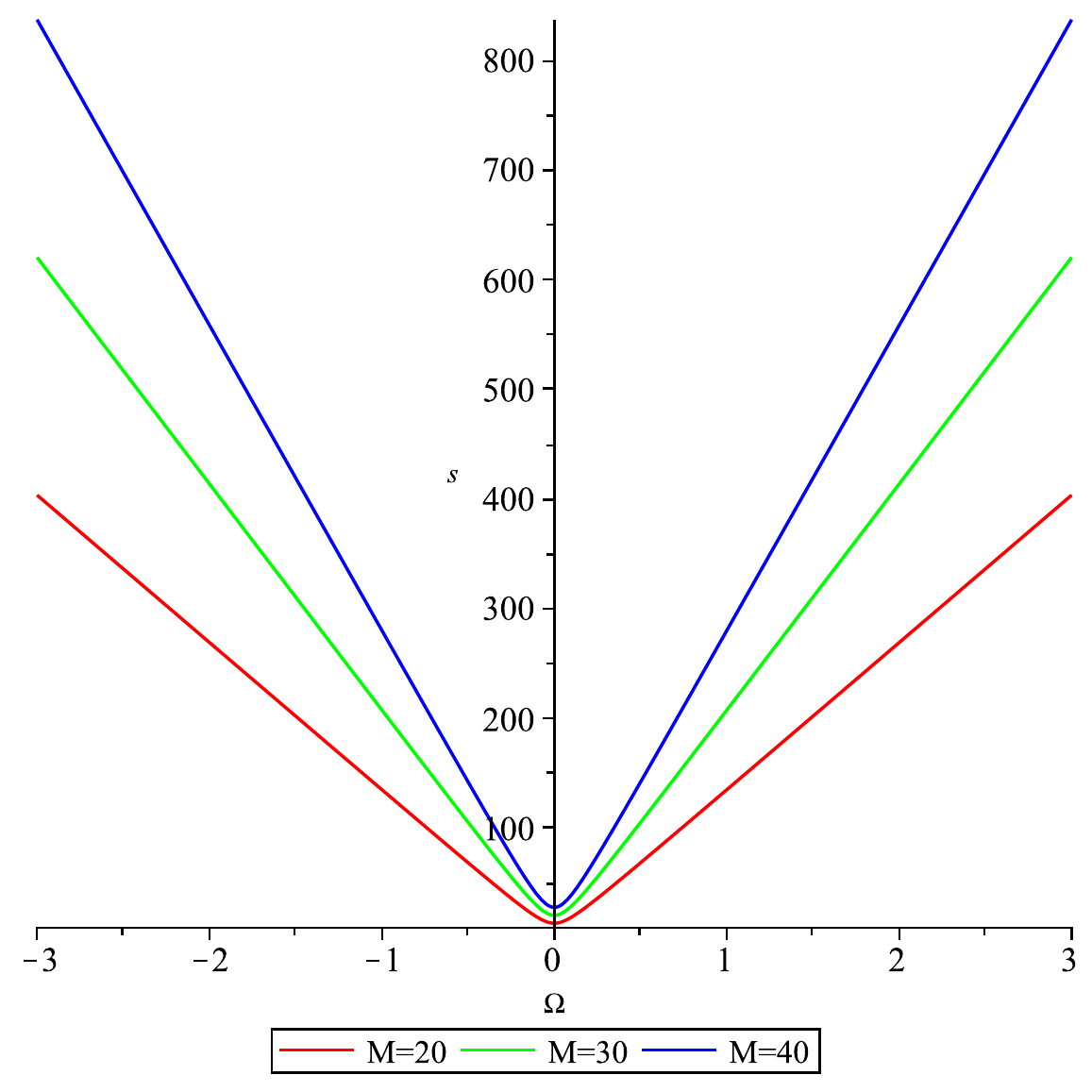}
		\caption{\centering} 
		\label{fig:2d}
	\end{subfigure}
	\caption{ (a) The entropy $S$ in terms of $M$ for three different charges, where we set $\Omega=0.1$, $l=0.1$, $\alpha=0.04$, $r_+=0.1$ and $L=1$. (b) The entropy $S$ in terms of $\Omega$ for three different masses, where we set $Q=0.1$, $l=0.1$, $\alpha=0.04$, $r_+=0.1$ and $L=1$.}
	\label{fig:ent}
\end{figure}

\begin{figure}[H]
\centering
\includegraphics[scale=0.50]{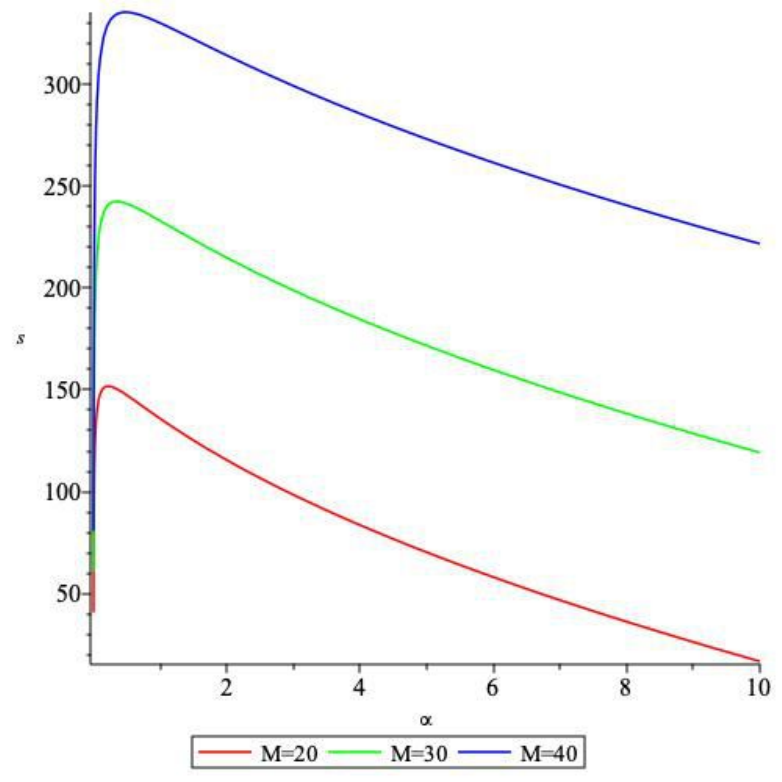}
\caption{{\textcolor{black}{The entropy $S$ in terms of $\vert \alpha\vert$ for three different masses, where we set $\Omega=0.1$, $l=0.1$, $r_+=0.1$, $Q=0.1$ and $L=1$.} }}
\label{fig:hab}
\end{figure}

The entropy of the black hole in $f(T)$ gravity, where $f''(T)<<1$, can be calculated as \cite{OLD4, OLD15}
\begin{equation}
    S_{BH}=f'(T) \frac{\mathcal{A}}{4},\label{SS}
\end{equation}
where prime denotes the derivation with respect to the scalar torsion $T$, and $\mathcal{A}$ is the area of the outer horizon. For the case where $f'(T)=1$, the entropy (\ref{SS}) becomes the Bekenstein-Hawking entropy.   We find $\mathcal{A}$ for the metric (\ref{lineel}) with $dr=dt=0$ as \cite{OLD4}
\begin{equation}
    \mathcal{A}=\int_0^{2\pi} \int_0^L d\phi dz \sqrt{-g}= \frac{2\pi r_+^2 \Xi L}{l},
\end{equation}
{\textcolor{black}{where $r_+$, the outer horizon for the black hole (\ref{lineel}), is the largest non-zero root of the $A(r)$, which is given by the algebraic equation
\begin{equation}
\frac{r_+^6}{6\vert \alpha \vert}+9Q^2r_+^4-6Mr_+^3+2Q^3\sqrt{6\vert\alpha\vert}=0.
\end{equation}
if $\alpha \neq 0$. This fact is in agreement with previous note that the black hole solution (\ref{lineel}) exists only for non-zero values of the $f(T)$-parameter $\alpha$. In other words, the black hole solution (\ref{lineel}) does not have any corresponding  GR limit.}}
Therefore, the entropy of the black hole is given by
\begin{equation}
    S=\frac{\pi \Xi L(7r_+^6+9r_+^4Q\sqrt{6|\alpha|}+18r_+^3M|\alpha|-54r_+^2Q^2|\alpha|-42Q^3\sqrt{6|\alpha|^3})}{9l(Q\sqrt{6|\alpha|}+r_+^2)^2}. \label{Sbh}
\end{equation}
{\textcolor{black}{as long as $\alpha \neq 0$}.} 
{\textcolor{black}{In Figure (\ref{fig:ent}) and (\ref{fig:hab}), we show the behaviour of the entropy in terms of the black hole parameters $Q$, $\Omega$ and the $f(T)$-parameter $\vert \alpha\vert$.}}
{\textcolor{black}{We notice the monotonic increasing behaviour of the black hole entropy versus the mass, as well as the angular velocity of the black hole. The increasing behaviours are in accord with the second law of thermodynamics for the black holes. We also notice that for very small values of $\vert \alpha \vert$, the entropy increases rapidly and reaches to maximum at about $\vert\alpha\vert \simeq 0.5$. Afterwards, the entropy decreases smoothly with increasing the values of $\vert \alpha \vert$. 
}}

\begin{figure}[H]
	\centering
	\begin{subfigure}{0.3\linewidth}
		\includegraphics[width=\linewidth]{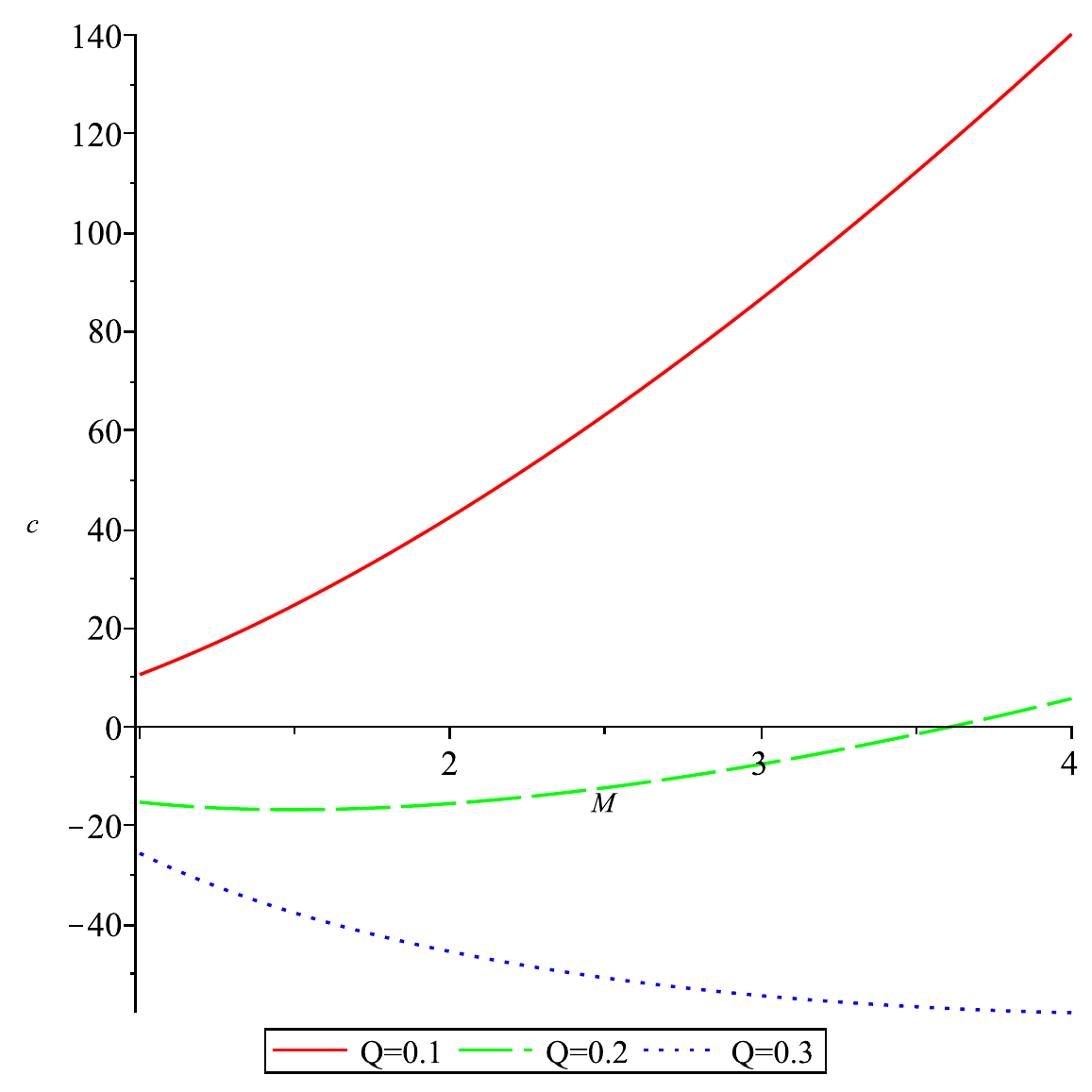}
		\caption{\centering}
		\label{fig:2a}
	\end{subfigure}
		\begin{subfigure}{0.3\linewidth}
		\includegraphics[width=\linewidth]{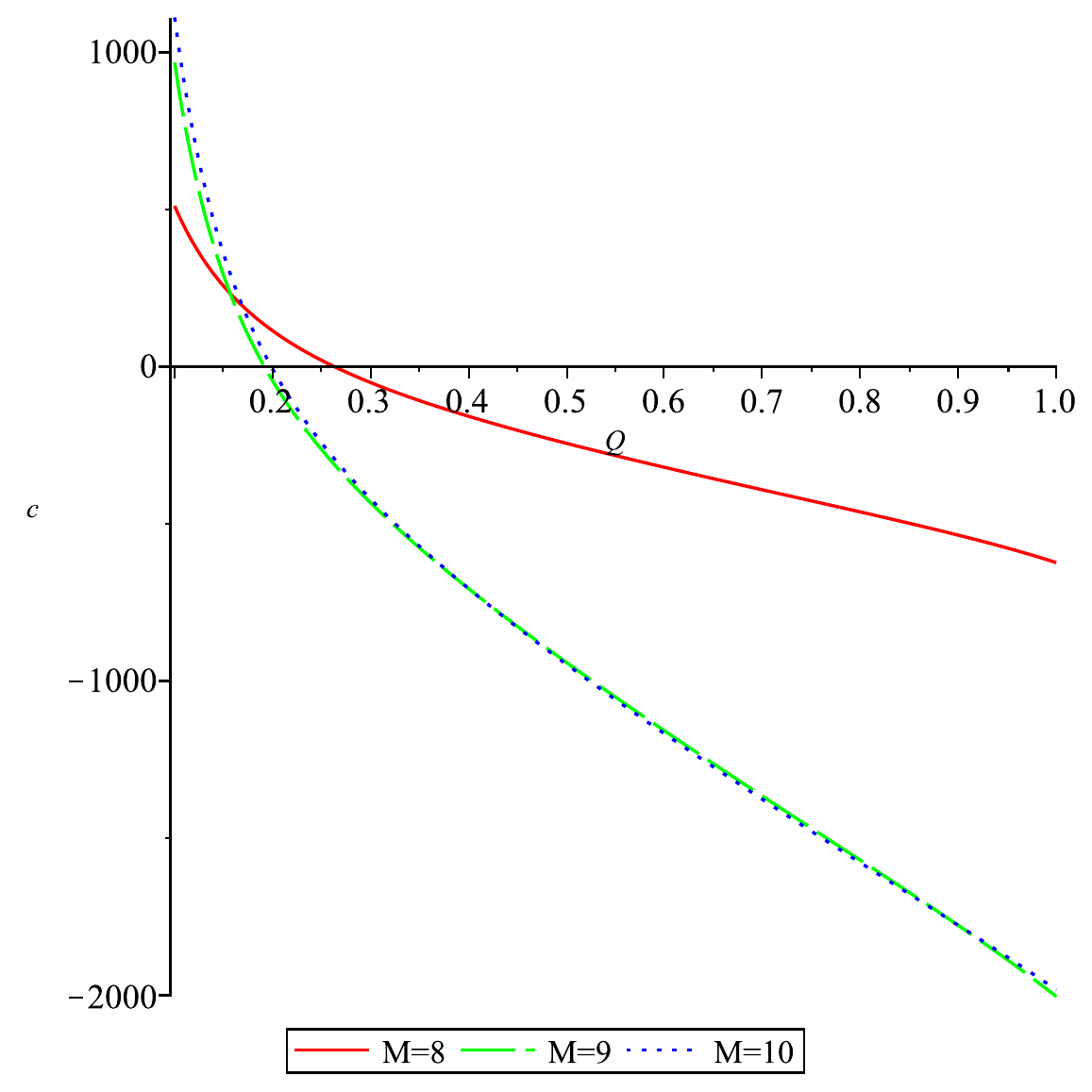}
		\caption{\centering}
		\label{fig:2a}
	\end{subfigure}
	\begin{subfigure}{0.3\linewidth}
		\includegraphics[width=\linewidth]{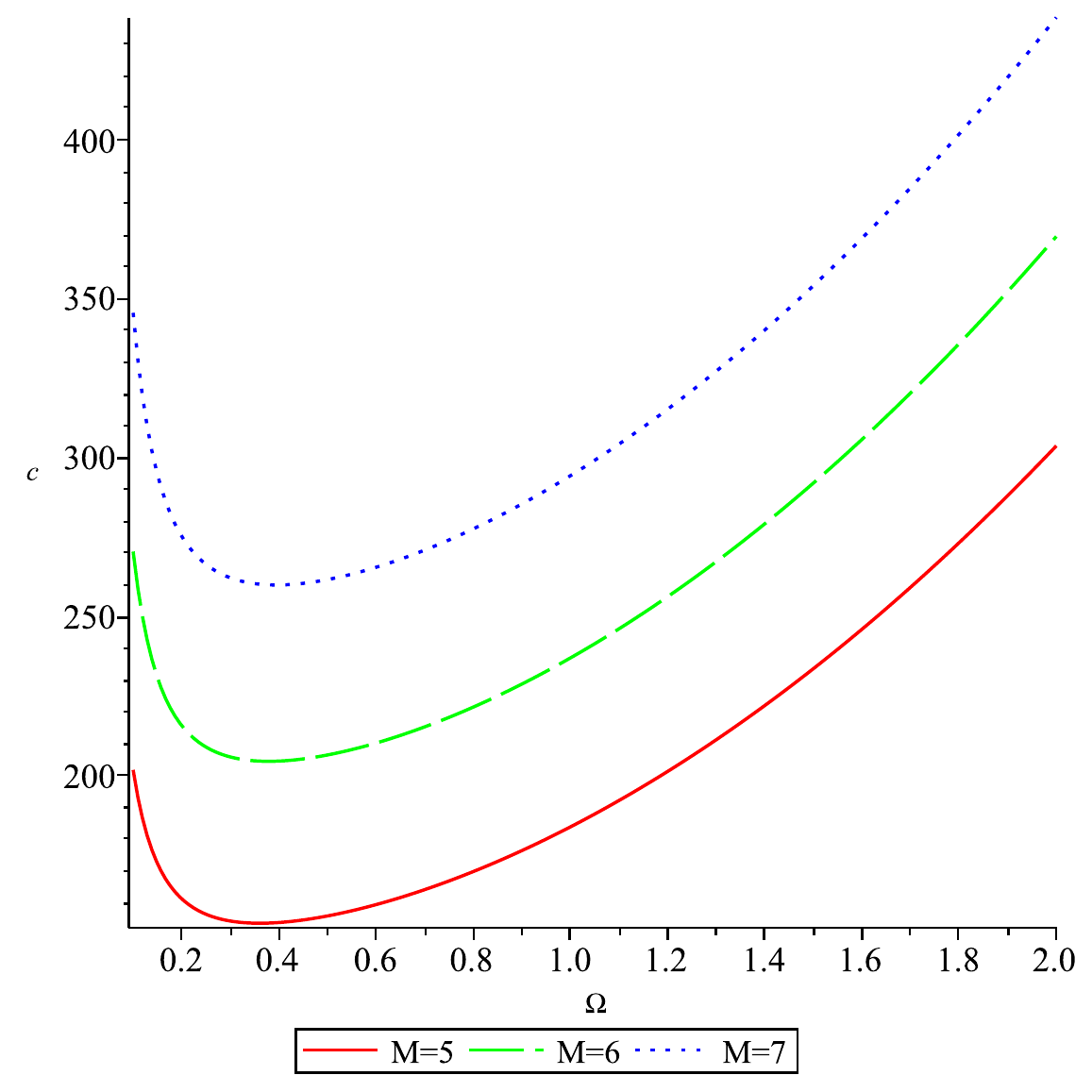}
		\caption{\centering}
		\label{fig:2b}
	\end{subfigure}
	\caption{The central charge in $J$ picture in terms of (a) $M$, (b) $Q$ and (c) $\Omega$, where we set $l=0.1$, $\alpha=0.004$, $r_+=0.1$, $r_*=0.2$ and $L=1$.}
	\label{fig: CJ}
\end{figure}

\begin{figure}[H]
	\centering
	\begin{subfigure}{0.3\linewidth}
		\includegraphics[width=\linewidth]{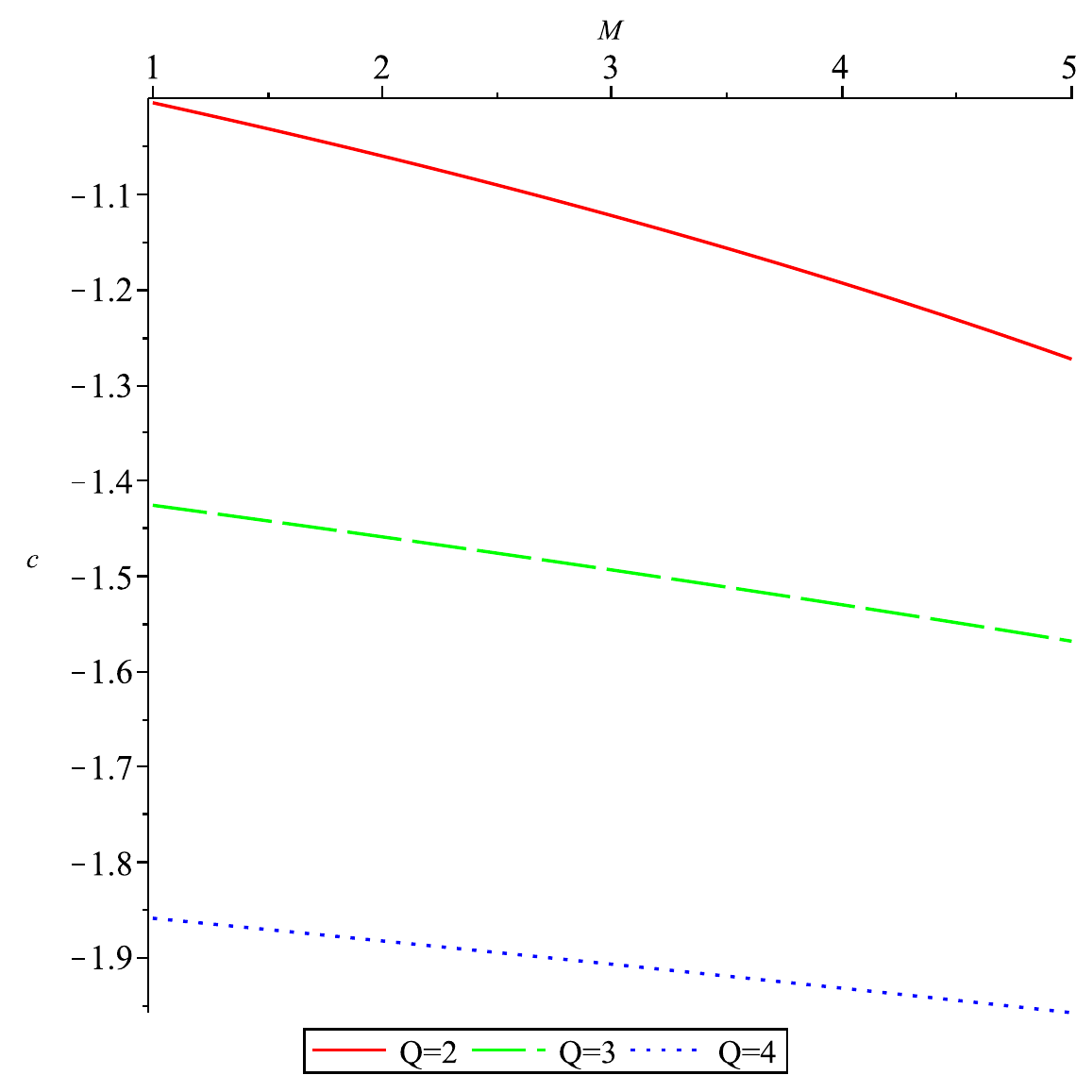}
		\caption{\centering}
		\label{fig:2c}
	\end{subfigure}
		\begin{subfigure}{0.3\linewidth}
		\includegraphics[width=\linewidth]{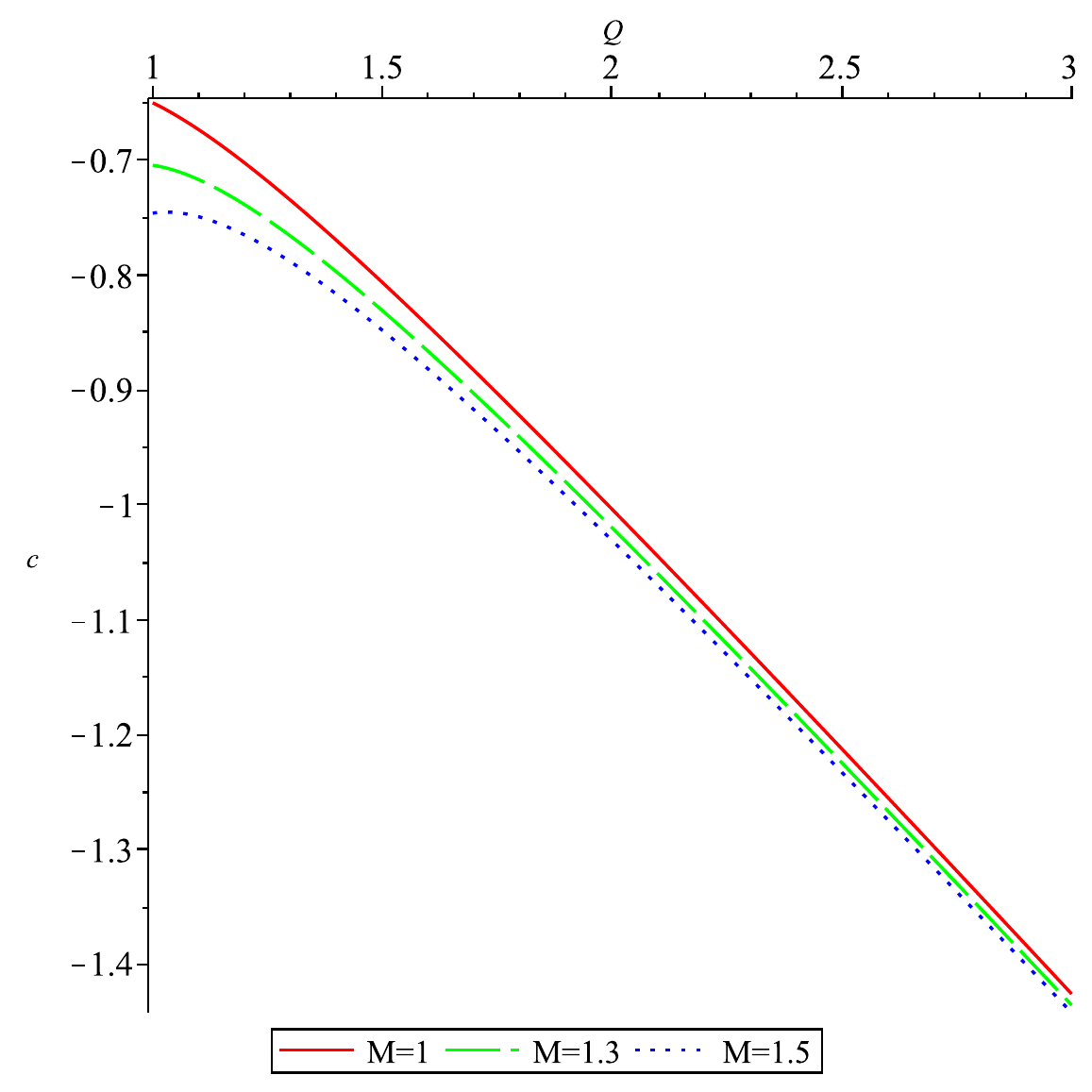}
		\caption{\centering}
		\label{fig:2a}
	\end{subfigure}
	\begin{subfigure}{0.3\linewidth}
		\includegraphics[width=\linewidth]{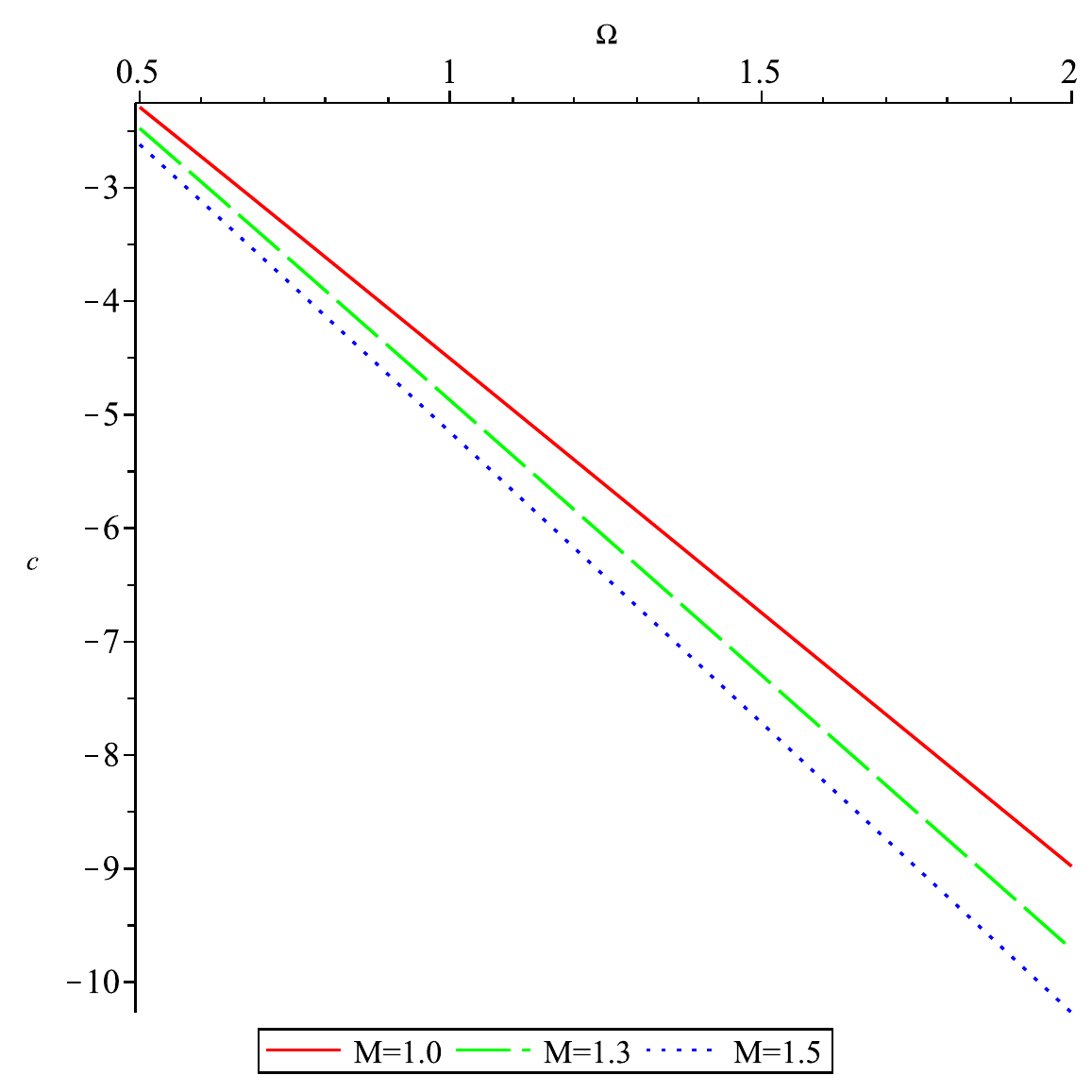}
		\caption{\centering} 
		\label{fig:2d}
	\end{subfigure}
	\caption{The central charge in $Q$ picture in terms of (a) $M$, (b) $Q$ and (c) $\Omega$, where we set $l=0.1$, $\alpha=0.004$, $r_+=0.1$, $r_*=0.2$ and $L=1$.}
	\label{fig: CQ}
\end{figure}

Comparing the Cardy formula for the entropy (\ref{cardy}) and the black hole entropy (\ref{Sbh}), as  well as $c_L=c_R$ \cite{OLD16}, we find the correspondence central charges for  $J$ picture as
\begin{equation}
    c^J_{L,R}=\frac{3}{\pi^2(T^J_L+T^J_R)}\frac{\pi \Xi L(7r_+^6+9r_+^4Q\sqrt{6|\alpha|}+18r_+^3M|\alpha|-54r_+^2Q^2\alpha-42Q^3\sqrt{6|\alpha|^3})}{9l(Q\sqrt{6|\alpha|}+r_+^2)^2},
\end{equation}
where $T^J_L$ and $T^J_R$ are given in (\ref{TJL}) and (\ref{TJR}), respectively.  For the $Q$ picture, we find
\begin{equation}
    c^Q_{L,R}=\frac{-\sqrt{6}XV}{\pi Y}(\frac{\pi \Xi L(7r_+^6+9r_+^4Q\sqrt{6|\alpha|}+18r_+^3M|\alpha|-54r_+^2Q^2|\alpha|-42Q^3\sqrt{6|\alpha|^3})}{9l(Q\sqrt{6|\alpha|}+r_+^2)^2}),
\end{equation}
where $V$, $X$ and $Y$ are given in (\ref{V}), (\ref{X}) and (\ref{Y}). 
Figures (\ref{fig: CJ}) and (\ref{fig: CQ}) show the behaviour of the central charge in terms of the black hole parameters $M$, $Q$ and $\Omega$ in $J$ and $Q$ picture, respectively.

{\textcolor{black}{We notice very different behaviour for the central charge (in the Planck units) versus the mass of black hole (in the Planck units) in $J$ and $Q$ picture, where we consider some arbitrary typical ranges for the black hole parameters and $\alpha$. In $J$ picture, the behaviour is more or less increasing with respect to the black hole mass, especially for low values of the electric charge. However in $Q$ picture, the central charge monotonically decreases with respect to mass of black hole, independent of the value of the electric charge. The central charge versus the charge of black hole, in both $J$ and $Q$ pictures, shows decreasing behaviour.  Moreover, we notice that the central charge in $J$ picture, initially decreases with the increasing the angular velocity of the black hole, reaches to a minimum around $\Omega \simeq 0.3$ and then increases monotonically. On the other hand, the central charge in $Q$ picture decreases monotonically with increasing the angular velocity of the black hole.
}}

\section{Hidden conformal symmetry for the extremal Charged Rotating Black holes in quadratic $f(T)$ gravity}
\label{sec:sec4:absorption}

In the present section, we consider the extremal four-dimensional charged rotating AdS black hole in {\textcolor{black}{quadratic}}  $f(T)$-Maxwell theory. The conformal coordinates in the extremal case are given below, which are different than the ones defined for the generic black hole 
\begin{equation}
    \omega^+=\frac{1}{2}(\alpha_1t+\beta_1\phi-\frac{\gamma_1}{r-r_+}), \label{tri}
\end{equation}
\begin{equation}
    \omega^-=\frac{1}{2}(e^{2\pi T_L\phi+2n_Lt}-\frac{2}{\gamma_1}),
\end{equation}
\begin{equation}
    y=\sqrt{\frac{\gamma_1}{2(r-r_+)}}e^{\pi T_L\phi+n_Lt}, \label{trf}
\end{equation}
where $\alpha_1$, $\beta_1$, $\gamma_1$ are constants. The radial equation for the scalar field is given by \cite{OLD11}
\begin{equation}
    \partial_{r}((r-r_+)^2\partial_{r})R(r)+(\mathcal{A}_1+\mathcal{B}_1+\mathcal{C}_1)R(r)=0,
\end{equation}
where $\mathcal{A}_1$ and $\mathcal{B}_1$ are given by
\begin{equation}
\mathcal{A}_1=A_1+qA_2+q^2A_3+\omega^2A_4+m^2A_5,
\end{equation}
where
\begin{equation}
    A_1=\frac{l^2(120Q\sqrt{6|\alpha|}r_+^2+72r_+^4+288Q^2|\alpha|)(-1/2Kk^2\Omega^4+Kk^2l^2\Xi^2\Omega^2+\omega m\Xi(Kl^2-6)\Omega-1/2Kk^2l^4\Xi^4)}{r_+^5(\Xi^2l^2-\Omega^2)^2K^2(r-r_+)},
\end{equation}
\begin{equation}
    A_2=\frac{36Q(10\sqrt{6|\alpha^3|}Q^3+21r_+^4\sqrt{6|\alpha|}Q+11r_+^6+70Q^2|\alpha|r_+^2)(-\Xi l^2\omega+\Omega m)}{r_+^8(-\Xi^2l^2+\Omega^2)K^2(r-r_+)},
\end{equation}
\begin{equation}
    A_3=\frac{144Q^2(8/3Q\sqrt{6|\alpha|}r_+^6+5/4r_+^8+8/3Q^3\sqrt{6|\alpha^3|}r_+^2+11Q^2|\alpha|r_+^4+Q^4\alpha^2)}{K^2(r-r_+)r_+^{11}},
\end{equation}
\begin{equation}
    A_4=-\frac{60l^4(K\Omega^2-6\Xi^2)(\sqrt{6|\alpha|}Qr_+^2+3/2r_+^4+12/5Q^2|\alpha|)}{r_+^5(\Xi^2l^2-\Omega^2)^2K^2(r-r_+)},
\end{equation}
\begin{equation}
    A_5=-\frac{60(K\Xi^2l^4-6\Omega^2)(\sqrt{6|\alpha|}Qr_+^2+3/2r_+^4+12/5Q^2|\alpha|)}{r_+^5(\Xi^2l^2-\Omega^2)^2K^2(r-r_+)},
\end{equation}
and
\begin{equation}
    \mathcal{B}_1=B_1+qB_2+q^2B_3+\omega^2B_4+m^2B_5,
\end{equation}
where
\begin{equation}
    B_1=-\frac{2l^2(Q\sqrt{6|\alpha|}+r_+^2)^2(-1/2Kk^2\Omega^4+Kk^2l^2\Xi^2\Omega^2+\omega m\Xi(Kl^2-6)\Omega-1/2Kk^2l^4\Xi^4)}{r_+^4(\Xi^2l^2-\Omega^2)^2K^2(r-r_+)^2},
\end{equation}
\begin{equation}
    B_2=-\frac{(Q\sqrt{6|\alpha|}+r_+^2)^24Q(-\Omega m+\Xi l^2\omega)(\sqrt{6|\alpha|}Q+3r_+^2)}{r_+^7K^2(r-r_+)^2(\Xi^2l^2-\Omega^2)},
\end{equation}
\begin{equation}
    B_3=-\frac{(Q\sqrt{6|\alpha|}+r_+^2)^24Q^2(\sqrt{6|\alpha|}Qr_+^2+3/2r_+^4+|\alpha| Q^2)}{K^2(r-r_+)^2r_+^{10}},
\end{equation}
\begin{equation}
    B_4=\frac{(Q\sqrt{6|\alpha|}+r_+^2)^2l^4(K\Omega^2-6\Xi^2)}{(r-r_+)^2(\Xi^2l^2-\Omega^2)^2K^2r_+^4},
\end{equation}
\begin{equation}
    B_5=\frac{(Q\sqrt{6|\alpha|}+r_+^2)^2(K\Xi^2l^4-6\Omega^2)}{(r-r_+)^2(\Xi^2l^2-\Omega^2)^2K^2r_+^4}.
\end{equation}
Moreover, we find $\mathcal{C}_1$ is given by replacing $r_*$ with $r_+$ in equation (\ref{C1}). 

We define the locally conformal operators as in equations (\ref{opi})-(\ref{opf}) based on the coordinate transformations (\ref{tri})-(\ref{trf}). We find the quadratic Casimir operators in terms of coordinates $(t,r,\phi)$ as \cite{OLD11}
\begin{equation}
    \mathcal{H}^2=\Tilde{ \mathcal{H}}^2=\partial_r((r-r_+)^2\partial_r)-(\frac{\gamma_1(2\pi T_L\partial_t-2n_L\partial_{\phi})}{\hat{G}(r-r_+)})^2-\frac{2\gamma_1(2\pi T_L\partial_t-2n_L\partial_{\phi})}{\hat{G}^2(r-r_+)}(\beta_1\partial_t-\alpha_1\partial_{\phi}),
\end{equation}
where $\hat{G}=2\pi T_L\alpha_1-2n_L\beta_1$. The radial equation in terms of $SL(2,\mathbb{R})\times SL(2,\mathbb{R})$ can be rewritten as
\begin{equation}
    \mathcal{H}^2R(r)=\Tilde{ \mathcal{H}}^2R(r)=-\mathcal{C}_1R(r). \label{e107}
\end{equation}
Assuming the transformation (\ref{e74}) and setting $q'=0$, we find the general $J'$ picture. 
We find the CFT parameters in $J'$ picture as
\begin{equation}
    n^{J'}_L=\sqrt{\frac{4\gamma_1\pi a_1}{\hat{\mathbf{A}} (2\pi b_1-2a_1)^2}},\label{nLJP}
\end{equation}
\begin{equation}
    \alpha_1=b_1.n^{J'}_L,
\end{equation}
\begin{equation}
    T^{J'}_L=\frac{2\gamma_1n_L^{J'}}{(2\pi\alpha_1-2a_1n_L^{J'})\sqrt{\hat{\mathbf{D}}}},
\end{equation}
\begin{equation}
    \beta_1=a_1T_L^{J'}
\end{equation}
\begin{equation}
    n^{J'}_R=0, T^{J'}_R=0,
\end{equation}
where the quantities $a_1$, $b_1$, $\hat{\mathbf{A}}$, $\hat{\mathbf{B}}$, $\hat{\mathbf{C}}$ and $\hat{\mathbf{D}}$ are given by
\begin{equation}
    a_1=\frac{\gamma_1\pi \hat{\mathbf{A}}}{\hat{\mathbf{C}}}
\end{equation}
\begin{equation}
    b_1=\frac{\gamma_1\hat{\mathbf{B}}}{\hat{\mathbf{D}}},
\end{equation}
\begin{equation}
    \hat{\mathbf{A}}=-\frac{60(K\Omega^2-6\Xi^2)(Q\sqrt{6|\alpha|}r_+^2+3/5r_+^4+12/5Q^2|\alpha|)l^4}{r_+^5(r-r_+)(\Xi^2l^2-\Omega^2)^2K^2},
\end{equation}
\begin{eqnarray}
    \hat{\mathbf{B}}&=&\frac{1}{r_+^{11}(\Xi^2l^2-\Omega^2)^2K^2}\{384Q^4\sqrt{6|\alpha|^3}r_+^2\lambda(\Xi^2l^2-\Omega^2)(-15/16\gamma\Omega r_++\lambda Q(\Xi^2l^2-\Omega^2)) \nonumber\\
    &-& 60Q\sqrt{6|\alpha|}r_+^6(\gamma^2(K\Xi^2l^4-6\Omega^2)r_+^2+63/5\lambda Q\Omega\gamma r_+(\Xi^2l^2-\Omega^2) \nonumber\\
    &-& 32/5Q^2(\Xi^2l^2-\Omega^2)^2)-36\gamma^2r_+^{10}(K\Xi^2l^4-6\Omega^2)+(-396\Omega\Xi^2\gamma l^2+396\gamma\Omega^3)\lambda Qr_+^9 \nonumber\\
    &+& 180\lambda^2Q^2(\Xi^2l^2-\Omega^2)^2r_+^8-144Q^2|\alpha|\gamma^2(K\Xi^2l^4-6\Omega^2)r_+^6-2520\lambda Q^3\Omega|\alpha|\gamma r_+^5 \nonumber\\
    &\times& (\Xi^2l^2-\Omega^2)+1584\lambda^2Q^4|\alpha| r_+^4(\Xi^2l^2-\Omega^2)^2+144\lambda^2Q^6\alpha^2(\Xi^2l^2-\Omega^2)^2\},
\end{eqnarray}
\begin{equation}
    \hat{\mathbf{C}}=\frac{(K\Omega^2l^4r_+^6-6\Xi^2l^4r_+^6)(Q\sqrt{6|\alpha|}+r_+^2)^2}{r_+^{10}K^2(-\Xi^2l^2+\Omega^2)^2},
\end{equation}
\begin{eqnarray}
    \hat{\mathbf{D}}&=&\frac{(Q\sqrt{6|\alpha|}+r_+^2)^2}{r_+^{10}K^2(\Xi^2l^2-\Omega^2)^2}\{-4Q^2r_+^2\lambda(-\gamma\Omega r_++\lambda Q(\Xi^2l^2-\Omega^2))(\Xi^2l^2-\Omega^2)\sqrt{6|\alpha|} \nonumber\\
    &+& \gamma^2r_+^6(K\Xi^2l^4-6\Omega^2)+(12\Omega\Xi^2\gamma l^2-12\Omega^3\gamma)\lambda Qr_+^5-6\lambda^2Q^2r_+^4(\Xi^2l^2-\Omega^2)^2 \nonumber\\
    &-& 4\lambda^2Q^4|\alpha|(\Xi^2l^2-\Omega^2)^2\}.
\end{eqnarray}
 We notice for unit element of  $SL(2,\mathbb{Z})$ (where $J'$ picture reduces to $J$ picture),  where $\gamma= 1$  and  $\lambda=0 $, we find $\pi b_1=a_1$. As a result, the CFT quantity  (\ref{nLJP}) in $J$ picture approaches infinity. This indicates that the $J$ picture is not well defined for the extremal rotating charges AdS black holes in {\textcolor{black}{quadratic}} $f(T)$ gravity. The absence of one picture  for  charged rotating black holes was previously discovered for the Kerr-Sen black holes \cite{OLD10}. It's quite an interesting research subject to address the existence or non-existence of one or two pictures in different theories of gravity.
 {\textcolor{black}{
 \begin{figure}[H]
	\centering
		\begin{subfigure}{0.33\linewidth}
		\includegraphics[width=\linewidth]{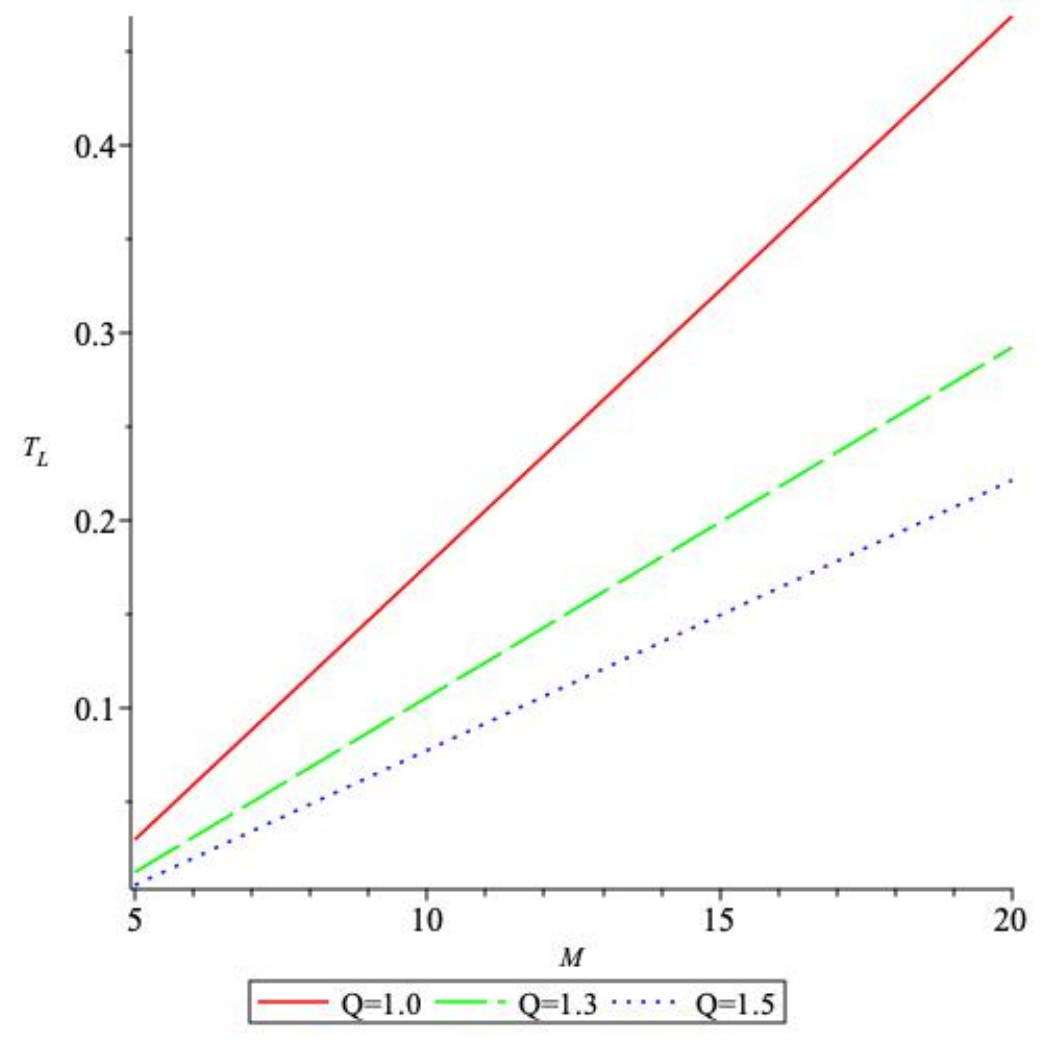}
		\caption{\centering}
		\label{fig:2c}
	\end{subfigure}
	\begin{subfigure}{0.33\linewidth}
		\includegraphics[width=\linewidth]{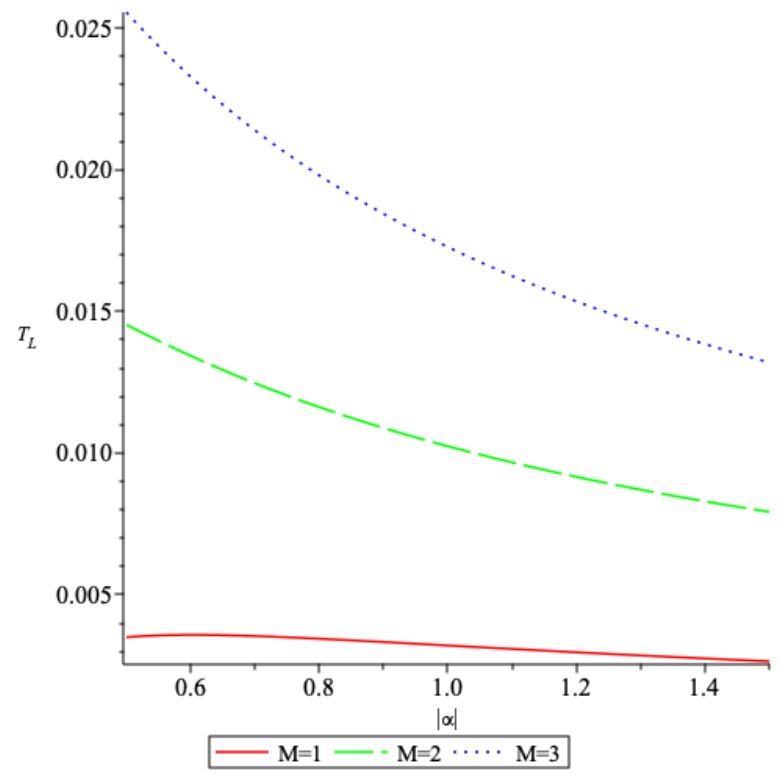}
		\caption{\centering} 
		\label{fig:2d}
	\end{subfigure}
	\caption{ {\textcolor{black}{The behaviour of $T_L$ in terms of $M$ and $\alpha$ for the extremal case in $Q$ picture, where we set $l=0.1$, $\Omega=5$.}}}
	\label{fig:ext1}
\end{figure}
\begin{figure}[H]
	\centering
		\begin{subfigure}{0.33\linewidth}
		\includegraphics[width=\linewidth]{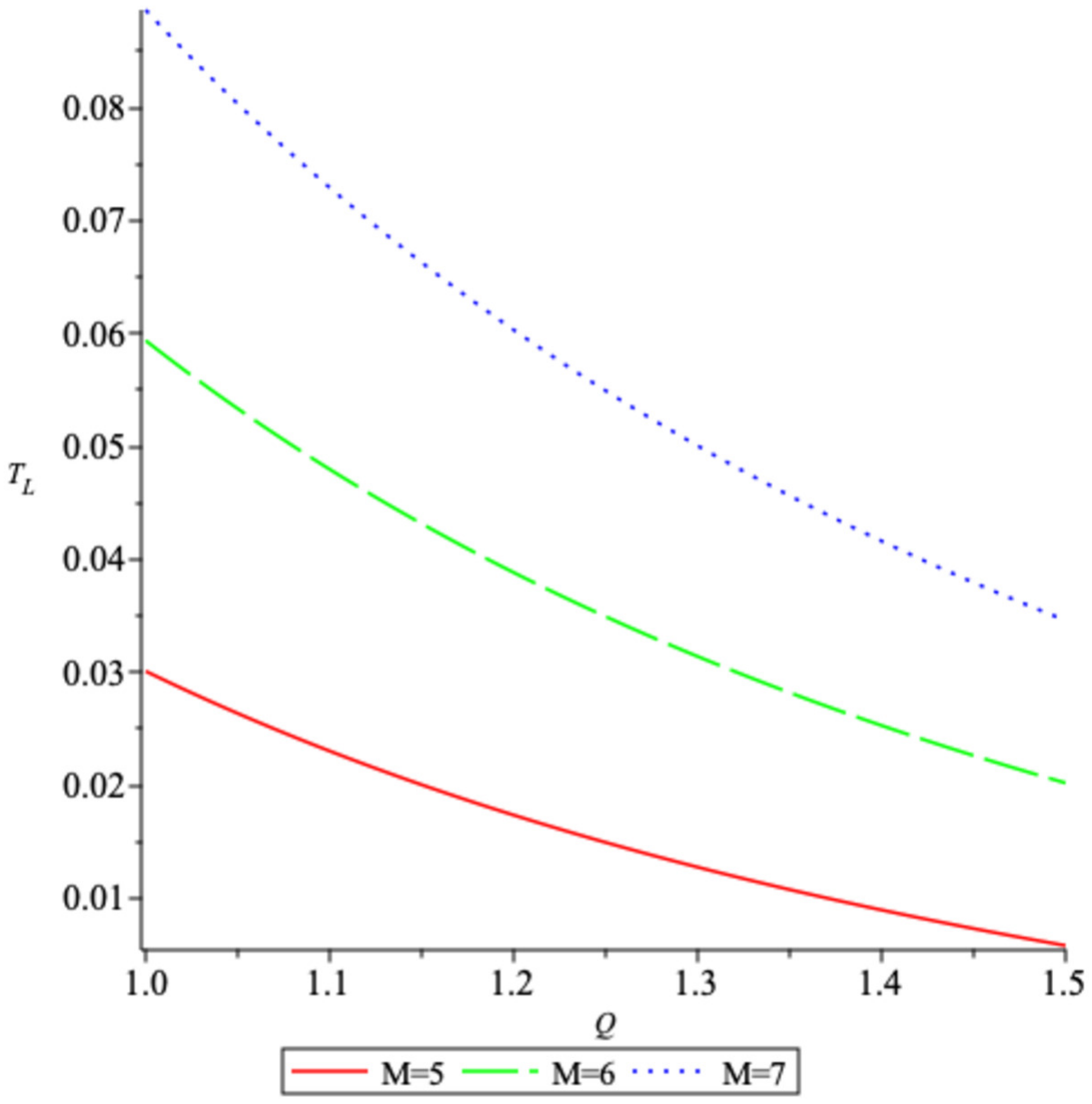}
		\caption{\centering}
		\label{fig:2c}
	\end{subfigure}
	\begin{subfigure}{0.33\linewidth}
		\includegraphics[width=\linewidth]{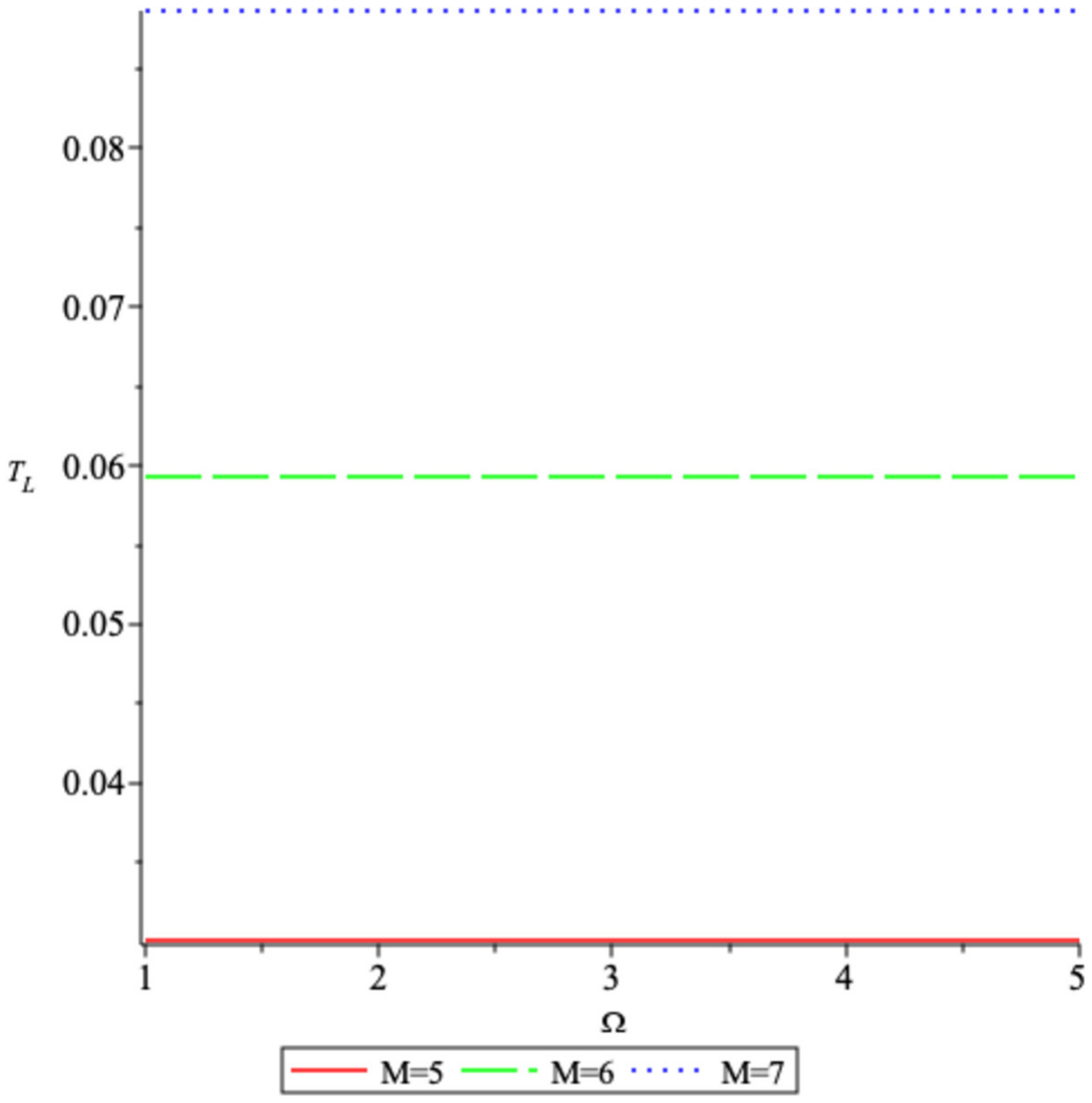}
		\caption{\centering} 
		\label{fig:2d}
	\end{subfigure}
	\caption{ {\textcolor{black}{The behaviour of $T_L$ in terms of $Q$ and $\Omega$ for the extremal case in $Q$ picture, where we set $l=0.1$ $|\alpha|=0.04$.}}}
	\label{fig:ext2}
\end{figure}
}}

The correspondence $Q'$ picture can be found by assuming $m'=0$ instead of $q'=0$. Moreover, for the unit element of  $SL(2,\mathbb{Z})$ where $\eta=0$ and $\tau=1$, the results of $Q'$ picture reduce to the  results for the $Q$ picture, in the extremal case. {\textcolor{black}{Figures (\ref{fig:ext1}) and (\ref{fig:ext2}) show the behaviour of $T_L$ (in the Planck units) in terms of the black hole parameters $M$, $Q$, $\Omega$ and $\alpha$ (all in the Planck units) in extremal $Q$ picture, where we consider some arbitrary typical ranges for the black hole parameters and $\alpha$. As we notice $T_L$ has a monotonically sharp increasing behaviour with respect to the mass of the black hole. On the other hand, $T_L$ is decreasing with respect to increasing the $f(T)$-parameter $\alpha$. The behaviour of $T_L$ versus the electric charge of the black hole shows a monotonically decreasing behaviour. Finally, $T_L$ is almost constant with respect to the angular velocity of the black hole. 
}}

\section{Conclusions}
\label{sec:concl}

{\textcolor{black}{In this paper, we generalize and extend the very well known black hole holography in GR, to the non-extremal and extremal four-dimensional charged rotating AdS black holes in the quadratic $f(T)$-Maxwell theory. }}

{\textcolor{black}{We explicitly construct three very different hidden conformal symmetries (known as $J$, $Q$ and general pictures) for the non-extremal charged rotating black holes in $f(T)$-Maxwell theory with a negative cosmological constant.  
We find that the charged scalar field in the background of charged rotating quadratic $f(T)$-Maxwell black holes, reveals the existence of three different CFTs.  We note that each CFT is dual to each black hole hairs, such as angular momentum and the electric charge.}}

{\textcolor{black}{To achieve the establishing the black hole holography, we mainly consider the near-horizon region, since the line element metric function, which determines the location of the event horizon, is a sixth order algebraic equation. In the near-horizon region, we show that the radial part of the field equation, for the charged scalar field, could be matched to the quadratic Casimir equation for an $SL(2,\mathbb{R})_L\times SL(2,\mathbb{R})_R$ conformal algebra. The existence of conformal algebra shows a local hidden conformal symmetry which acts on the solution space of the charged scalar field. }}

{\textcolor{black}{We also explicitly show that the $SL(2,\mathbb{R})_L\times SL(2,\mathbb{R})_R$ symmetry is spontaneously broken to $U(1)_L \times U(1)_R$ under the coordinate identification $\phi \sim \phi + 2\pi$. The  $U(1)_L \times U(1)_R$ subalgebra of the original conformal symmetry indicates that the charged rotating AdS black holes in quadratic $f(T)$ gravity, is dual to the finite temperatures $(T_L,T_R)$ mixed state in a two-dimensional CFT. }} 

{\textcolor{black}{We also calculate the central charges of the CFT, by matching the dual CFT Cardy entropy to the macroscopic Bekenstein-Hawking entropy of the black hole.  The results very clearly indicate that the 
charged rotating AdS black holes in quadratic $f(T)$-Maxwell theory with particular values of $M, \Omega, Q$, and $\alpha$, are indeed dual to different two-dimensional CFTs.}}

{\textcolor{black}{We then consider the class of extremal charged rotating AdS black holes in the quadratic $f(T)$ gravity. Using a new set of conformal coordinates, we match the radial part of the field equation, for the charged scalar field, to the quadratic Casimir invariant of the conformal symmetry. The results of the article , for the first time, show the validity and extension of the holographic duality for the charged rotating black holes beyond the GR black holes, and to the realm of the  $f(T)$ theories of gravity.}

{\textcolor{black}{We also find the interesting result that, unlike the non-extremal case, the only consistent dual CFT to the extremal black hole, is the $Q$ picture. This may indicate that  
we can't apply the “microscopic no hair conjecture”, proposed in \cite{NN} to the extremal charged rotating black holes of the quadratic $f(T)$ gravity. In fact, the reason is that the extremal charged rotating black holes of $f(T)$ gravity are not obtained, and also do not have any proper limit to approach the black hole solutions of the GR. The only limit that quadratic $f(T)$ black hole can reduce to a GR black hole is setting $\alpha=0$ in (\ref{fTa}), however, the metric functions (\ref{metra}) and (\ref{metrb}) are not well defined in this limit, as $\Lambda_{eff}\rightarrow \infty$.
}}

{\textcolor{black}{For future works, we plan to construct the extended family of the conformal symmetry for the charged rotating black holes of quadratic $f(T)$ gravity, in which the symmetry is deformed by a deformation parameter \cite{NN2}. We also plan to establish the existence of a super conformal field theory, with a global $U(1)$ symmetry, in which the $J$ and $Q$ pictures are related to the spectral flow transformations of the super conformal field theory. 
Moreover, finding the new charged rotating black holes in MTG theories, such as $f(R)$ and $f(Q)$ theories, not only provides a treasure trove of information about the rotating black holes in MTG, but also provides the opportunity to establish (or rule out) the very existence of the black hole holography in theories beyond the GR.  Such possibility of the establishments of the holography for rotating black holes beyond GR, can shed more lights on the nature of the black hole holography in the context of quantum gravity. 
}}

{\textcolor{black}{The other interesting open question is finding the near-horizon geometry of the near-extremal charged rotating black holes of the quadratic $f(T)$ gravity, as well as any other discovered new charged rotating black holes in MTG. Moreover finding the central charges of the dual CFT to the solutions of the MTG, using the asymptotic symmetry group, is another interesting open question. Though it is a very complicated task, however it provides another layer of confirmation on the duality between charged rotating black holes of MTG and CFTs, and especially our results in this article for the charged rotating black holes of quadratic $f(T)$ gravity.  The other interesting future work involves calculation of the different types of the super-radiant scattering, off the near-extremal black holes of MTG. The results definitely provide further evidence to support the holographic dual for the charged rotating black holes of MTG, as well as the charged rotating black holes of the quadratic $f(T)$ gravity.
}}

\vspace{1cm} 
\bigskip

{\Large Acknowledgments}\newline
This work was supported by the Natural Sciences and Engineering Research Council of Canada.\newline

\end{document}